\documentclass[11pt,a4paper]{article}
\pdfoutput=1

\usepackage{latexsym,amssymb,amsmath,amsfonts,amstext}
\usepackage{slashed}
\usepackage{mathrsfs}
\usepackage{bbold}
\usepackage{xcolor}
\usepackage[toc,page]{appendix}
\usepackage{jheppub}

\allowdisplaybreaks
\def\slash#1{\rlap{\hbox{$\mskip 1 mu /$}}#1}      	
\def\Slash#1{\rlap{\hbox{$\mskip 3 mu /$}}#1}      	

\title{One-loop determinants for black holes\\ in 4d gauged supergravity}

\author{Kiril Hristov$^a$,}
\author{Ivano Lodato$^b$,}
\author{and Valentin Reys$^{c,d}$}

\affiliation{$^a$ Institute for Nuclear Research and Nuclear Energy, \\ Bulgarian Academy of Sciences, 1784 Sofia, Bulgaria}
\affiliation{$^b$ Department of Physics and Center for Field Theory and Particle Physics, \\ Fudan University, 200433 Shanghai, China}
\affiliation{$^c$ Dipartimento di Fisica, Universit\`a di Milano--Bicocca, \\ Piazza della Scienza 3, I-20126 Milano, Italy}
\affiliation{$^d$ INFN, sezione di Milano--Bicocca, \\ Piazza della Scienza 3, I-20126 Milano, Italy}

\emailAdd{khristov@inrne.bas.bg, kionzivano@gmail.com, valentin.reys@unimib.it}

\abstract{We continue the effort of defining and evaluating the quantum entropy function for supersymmetric black holes in 4d ${\cal N} = 2$ gauged supergravity, initiated in \cite{Hristov:2018lod}. The emphasis here is on the missing steps in the previous localization analysis, mainly dealing with one-loop determinants for abelian vector multiplets and hypermultiplets on the non-compact space $\mathbb{H}_2 \times \Sigma_{\rm g}$ with particular boundary conditions. We use several different techniques to arrive at consistent results, which have a most direct bearing on the logarithmic correction terms to the Bekenstein-Hawking entropy of said black holes.}

\begin{document}

\maketitle


\section{Introduction}
\setcounter{equation}{0}

The topic of black holes in gauged supergravity has enjoyed a substantial attention in the literature in recent years, sparked by the successful holographic understanding of the leading microscopic degrees of freedom for a class of supersymmetric asymptotically AdS$_4$ solutions \cite{Benini:2015eyy}. A multitude of generalizations and related progress have since been reported (see \cite{Zaffaroni:2019dhb} for an overview of the subject). In the present paper, which can be regarded as a sequel to \cite{Hristov:2018lod}, we aim to address systematically the problem of finding the quantum corrections to the Bekenstein-Hawking formula for the entropy of black holes. In the pursuit of finding an exact macroscopic entropy formula for black holes in gauged supergravity, we used the approach initiated in \cite{Dabholkar:2010uh,Dabholkar:2011ec} for localization in supergravity. The steps taken in \cite{Hristov:2018lod} allowed us to determine the localization locus and classical action as will be reviewed shortly. The present work extends these steps to include the semi-classical analysis of one-loop determinants encoding the quadratic fluctuations around the localization locus.

The computation of one-loop determinants in supersymmetric localization (starting from the seminal work of \cite{Pestun:2007rz}, see \cite{Pestun:2016zxk} for a more pedagogical review) is in itself an already well-developed topic where various mathematical tools have been employed. It is worth stressing that the established mathematical theorems behind such calculations rely heavily on the assumption that the underlying space on which the quantum fields propagate is compact. We are instead interested in the quantum entropy function (QEF) \cite{Sen:2008vm} defined on the (Euclidean) near-horizon solution of the supersymmetric (BPS) black holes of interest, 
\begin{equation}
\label{eq:first-macroscopic-degeneracy}
d_{\rm macro} (p^I, q_I) := \left< \exp \left( 4\pi\,q_I \int_0^{2\pi} W^I_{\tau}\,{\rm d} \tau \right) \right>_{\mathrm{EAdS}_2}^{\mathrm{finite}}\ ,
\end{equation}
where the Wilson line insertion enforces the microcanonical ensemble with fixed electric charges (as opposed to the canonical ensemble of fixed chemical potentials). We work in four dimensions and thus the underlying space we are lead to consider is the non-compact space $\mathbb{H}_2 \times S^2$, where $\mathbb{H}_2$ is the hyperbolic disk (i.e. Euclidean AdS$_2$), and we can further replace the $S^2$ factor by an arbitrary genus Riemann surface $\Sigma_{\rm g}$. We will nevertheless go ahead and often use a given theorem under the extra assumption that for our purposes the statement of compactness can be replaced by a careful choice of boundary conditions, which are known to be crucial at the asymptotic boundary of Euclidean AdS spaces. A similar approach has already been advocated and successfully used for the computation of one-loop determinants in a number of interesting examples \cite{Murthy:2015yfa,Gupta:2015gga,David:2016onq,Assel:2016pgi,Cabo-Bizet:2017jsl,David:2018pex,David:2019ocd}.

We focus our one-loop analysis mostly on the contributions from an arbitrary number $n_V$ of abelian vector multiplets and hypermultiplets (both the compensating one and a possible number $n_H$ of physical ones) in the conformal supergravity formalism. In this context we should note that the results we present here have a stand-alone character if one is also interested in localization of rigid 4d ${\cal N}=2$ field theories on $\mathbb{H}_2 \times \Sigma_{\rm g}$, in which case the multiplets we consider are the main constituents of many gauge theories coupled to matter. Our interest here is however stemming from the bulk black hole physics and consequently we need to supplement the aforementioned one-loop contributions with those coming from the gravitational degrees of freedom. We therefore also comment without rigorous derivation on contributions from the off-shell gravity Weyl multiplet as well as Kaluza-Klein modes, i.e.\ massive multiplets, that have been truncated away in the effective 4d description but need to be taken into account in a full 10d/11d calculation. 

Apart from their general importance for the evaluation of the QEF, to which we come back at the end of this paper, our results for the one-loop determinants of the various supergravity multiplets also have a more direct interpretation as logarithmic corrections to the leading Bekenstein-Hawking entropy. In this sense they are related to the logarithmic corrections to the entropy of asymptotically AdS$_4$ black holes derived in \cite{Liu:2017vll,Jeon:2017aif,Liu:2017vbl} using methods developed in \cite{Banerjee:2010qc,Banerjee:2011jp,Sen:2011ba,Bhattacharyya:2012ye}, as well as to the field theoretic $\log N$ corrections to the leading large $N$ expression calculated numerically in \cite{Liu:2017vll,Liu:2018bac,PandoZayas:2019hdb}. Our results however concern off-shell supergravity and should be interpreted in a slightly different context since the localization procedure requires some further input, such as the localization measure together with the aforementioned gravity and massive multiplet contributions, that is still missing. Therefore a direct comparison with other results in the literature is challenging at this point, although we already observe an interesting structure for the $\log$-corrections. 

One of the most salient qualitative conclusions from our investigations is the following {\it large charge} expansion of the entropy
\begin{equation}
\label{eq:first-expansion}
	\log d_{\rm macro}(p^I,q_I) = \frac{A_H(p^I,q_I)}{4\, G_N} + k_1 \log \left( \frac{A_H(p^I,q_I)}{4\, G_N} \right) +k_2 \log \left( \frac{L^2_{AdS_2}(p^I,q_I)}{G_N}\right) +  \ldots \ .
\end{equation}
Here, $A_H$ is the area of the black hole horizon which can have the topology of any genus Riemann surface $\Sigma_{\rm g}$, $G_N$ is the Newton constant in four dimensions, $L_{AdS_2}$ is the length scale of the AdS$_2$ space on which the quantum entropy function is defined, $k_{1, 2}$ are constants, and the dots denote subleading terms. The important feature is that the expansion above is not only in terms of the black hole area, but also in terms of the length scale of AdS$_2$. While the two quantities are proportional for asymptotically flat black holes, this is not in general the case for black holes in gauged supergravity. Furthermore, we find that the one-loop determinant contribution to the QEF that we compute in the present work only contributes to the value of $k_2$. The value of $k_1$ is instead fixed by the correction to the saddle-point evaluation giving the leading Bekenstein-Hawking term above, 
\begin{equation}\label{k1}
	k_1 = - \frac12 (n_V - n_H)\ ,
\end{equation}
where holographically $(n_V - n_H)$ translates into the number of abelian flavor symmetries, i.e.\ the total number of global $U(1)$'s without counting the R-symmetry. Within the framework of the QEF, we are able to identify $k_1$ as the correction due to the the conversion of the partition function between the fixed chemical potential (canonical) ensemble and the fixed electric charge (microcanonical) ensemble, as anticipated from field theory in \cite{Benini:2016rke,Hosseini:2018qsx}.
We are unable to fully determine the value of $k_2$ at the present stage, but our one-loop computations lead us to the expression
\begin{equation}\label{k2}
	k_2 = (1-{\rm g})\Bigl(\frac14 (n_V + 1 - n_H) + a_0\Bigr)\ ,
\end{equation}
with ${\rm g}$ genus of the horizon, and we parametrize the as-of-yet unknown contribution of the additional gravity and massive multiplets by $a_0$. On the field theory side, $k_2$ can be directly evaluated in the canonical ensemble as done in \cite{Liu:2017vll,Liu:2018bac}.

The above expansion follows from the quantum entropy function formalism and therefore naturally makes use of the scales of AdS$_2$ and the internal space. For the purposes of microstate counting in an asymptotically AdS$_4$ black hole spacetime, however, one needs to make contact with the length scale of the asymptotic space in order to use the holographic dictionary. This suggests that for the holographically dual calculation, the {\it large $N$} expansion of the entropy can be read-off from
\begin{equation}
\label{eq:second-expansion}
	\log d_{\rm macro}(p^I,q_I) = \frac{A_H(p^I,q_I)}{4\, G_N} + ( k_1+k_2) \log \left( \frac{L^2_{AdS_4}}{G_N}\right) +  \ldots \ ,
\end{equation}
where we have used that both length scales in \eqref{eq:first-expansion} are proportional to $L^2_{AdS_4}$ up to constant factors in $N$ that only contribute to further subleading terms in \eqref{eq:second-expansion}. Thus, although we will be able to distinguish between different terms contributing separately to $k_1$ or $k_2$, the logarithmic corrections can be put together if one is interested in matching holographically with field theory computations.

The outline of the paper is as follows. In the rest of this section we first review some details of the previous localization steps that we need in order to proceed with the one-loop calculation. We then outline the main idea behind calculating the one-loop contribution using the fact that supersymmetry ensures a large cancellation between bosonic and fermionic modes. In section \ref{sec:vectors} we derive the one-loop determinant for vector multiplets, starting from the gauge-fixing procedure and proceeding to derive the main result using three distinct methods. We then regularize the result via zeta-function regularization and generalize it to include the case of arbitrary higher genus Riemann surface horizon topology. In section \ref{sec:hypers} we derive the one-loop contribution from the compensating hypermultiplet, as well as for other possible physical hypermultiplets in the conformal supergravity formalism. Finally, in section \ref{sec:final} we try to put together a more complete picture of the quantum entropy function after assuming certain behavior of the localization measure that we cannot yet derive from first principles.

\subsection{Review of previous localization steps}
\label{subsec:review}

Here we review the set-up used in \cite{Hristov:2018lod} to localize the path-integral \eqref{eq:first-macroscopic-degeneracy} defined in \cite{Sen:2008vm} and describe the quantum entropy of black hole solutions in 4d $\mathcal{N}=2$ gauged supergravity with near-horizon geometry $\mathbb{H}_2 \times S^2$. We choose coordinates
\begin{equation}
\label{eq:NHG-metric}
\mathrm{d}s^2 = \mathring{g}_{\mu\nu}\,\mathrm{d}x^\mu\,\mathrm{d}x^\nu = v_1\,\bigl(\sinh^2\eta\,\mathrm{d}\tau^2 + \mathrm{d}\eta^2\bigr) + v_2\,\bigl(\mathrm{d}\theta^2 + \sin^2\theta\,\mathrm{d}\varphi\bigr) \, ,
\end{equation} 
and turn on the relevant gauge and auxiliary fields in the Weyl multiplet to support half-BPS solutions, as well as a set of electric and magnetic charges and scalar fields in the vector multiplets fixed on-shell by the attractor mechanism of gauged supergravity. The complete half-BPS near-horizon solution was derived in \cite{deWit:2011gk,Hristov:2016vbm} in the superconformal formalism and rewritten in \cite{Hristov:2018lod} in the Euclidean formulation of supergravity \cite{deWit:2017cle}. We gather the essential features of the near-horizon solution in appendix \ref{app:A}. There we also give the explicit form of the Killing spinors that were used for localization, given by a doublet of commuting symplectic-Majorana spinors $(\xi^i, \kappa^i)$ corresponding to supersymmetry and conformal supersymmetry parameters for the localizing supercharge $Q$.

We will discuss in detail the $Q$-exact deformation used to localize the path-integral that follows from our choice of localizing supercharge for vector multiplets and hypermultiplets in the dedicated sections. For now it is important to recall the corresponding algebra,
\begin{equation}
\label{eq:Q-alg}
Q^2 = \mathcal{L}_{\mathring{v}} + \delta_{\mathrm{R}}(\Lambda^i{}_j) + \delta_{\mathrm{gauge}}(\Lambda^I) \, , 
\end{equation}
where the parameters are as follows:
\begin{align}
\label{eq:KV}
\mathring{v}^\mu :=&\; -2\mathrm{i}\,\bar{\xi}_{i+}\gamma^\mu\xi^i_- = \frac{2}{\sqrt{v_1}}\,\bigl(1\,,0\,,0\,,0\bigr)^\mathrm{T} \, , \\[1mm]
\label{eq:SU2-param}
\Lambda^i{}_j :=&\; -2\,\bar{\xi}_j\gamma^5\kappa^i + \delta^i{}_j\,\bar{\xi}_k\gamma^5\kappa^k = \tfrac{1}{\sqrt{v_1}}\,\mathrm{i}\,\sigma_3{}^i{}_j \\[1mm]
\label{eq:Lambda-param}
\Lambda^I :=&\; 2\,\bar{\xi}_{i-}\xi^i_-\,X_+^I - 2\,\bar{\xi}_{i+}\xi^i_+\,X_-^I - \mathring{v}^\mu W_\mu^I \\
=&\; 2\,(\cosh\eta + 1)X_+^I - 2\,(\cosh\eta - 1)X_-^I - 2\sinh\eta\,W_1^I \nonumber \, .
\end{align}
Observe that the gauge transformation in \eqref{eq:Q-alg} depends explicitly on the  the gauge fields $W^I_{\mu}$ and scalars $X^I_{\pm}$ in the vector multiplets. Since we are ultimately evaluating a path-integral, we need to first gauge-fix the action and introduce the corresponding ghost fields, as will be discussed in due course. Another important observation is that the action of $Q^2$ on the near-horizon spacetime has a fixed ``point'' on a codimension two submanifold, namely the full $S^2$ (or in general $\Sigma_{\rm g}$) sitting at the centre of $\mathbb{H}_2$. This will play an important role in the evaluation of the one-loop determinants. 

Let us also briefly describe the result for the localization locus derived in \cite{Hristov:2018lod}. It was found that the vector multiplet fields are allowed to fluctuate away from their on-shell value in a specific way depending on $n_V$ real parameters $\phi^I_+$ and a set of $n_V + 1$ functions on the sphere $\phi^I_- (\theta, \varphi)$. The localized path-integral therefore remains infinite-dimensional, but due to the fact that the classical action was found to depend only on $\phi_+$, we could already write the resulting quantum entropy function in the following suggestive form\footnote{Here we write the result as an integral over $(\phi_+^I,\phi_-^I)$ instead of the set of fields $(\phi_+^I,\phi_0^I)$ originally used in \cite{Hristov:2018lod}. The reason for this change will be clear from our one-loop determinant computation after we come back to the QEF in section \ref{sec:final}.},
\begin{align}
\begin{split}
\label{eq:dmacro-loc-diverge}
&d_{\rm macro}(p^I,q_I) = \\
&\quad \int_{-\infty}^{+\infty} \,\Bigl(\prod_{I=0}^{n_V}\,\mathrm{d}\phi_+^I\Bigr)\;\delta\Bigl(\xi_I\phi^I_+ - \frac{1}{2\sqrt{v_1}}\Bigr)\,e^{-S_{\rm cl}[p^I, q_I, \phi^I_+]} \int\,\Bigl(\prod_{I=0}^{n_V}\,\mathcal{D}\phi_-^I\Bigr)\,Z_{\rm ind}(\phi_+, \phi_-) \, . 
\end{split}
\end{align}
This form allowed us to show that the saddle-point approximation, at $\mathring{\phi}_+^I = 2\mathring{X}^I_+$ where $\mathring{X}^I_+$ are the attractor values of the scalar fields, agrees with the expected classical entropy function $S_{\rm cl}[p^I, q_I, \mathring{\phi}_+^I]$ used for the leading-order holographic entropy matching \cite{Benini:2015eyy,Benini:2016rke}. In the present paper, we are after the remaining piece in \eqref{eq:dmacro-loc-diverge}, which further splits into
\begin{equation}
Z_{\rm ind}(\phi_+, \phi_-) = Z_{\text{1-loop}} (\phi_+, \phi_-)\ Z_{\rm measure} (\phi_+, \phi_-) \, . 
\end{equation}
We will be able to evaluate rigorously $Z_{\text{1-loop}}$ for vector multiplets and hypermultiplets and parametrize the contribution from the gravitational and massive multiplets. The last remaining factor, the localization measure, remains the least understood part of the supergravity localization formalism. We will briefly comment on it at the end of this paper.

\subsection{The one-loop set-up}
\label{subsec:looprev}

Let us now concentrate on the main subject of the present work, the computation of the one-loop determinant(s). We are specifically interested in the determinant of the quadratic fluctuations around the localization term added to deform the classical action, the operator $\widehat{Q} \widehat{\mathcal{V}}$. Here, hats remind us that we are dealing with a gauge-fixed theory, so some preliminary steps need to be taken when dealing with the localization term. For the moment we proceed abstractly in order to outline the main logic of the procedure we follow and come back to the explicit calculations in the bulk of the paper. 

The fermionic deformation $\widehat{\mathcal{V}}$ used for localization, expanded to quadratic order in the fields, can be written as  follows \cite{Pestun:2007rz}:
\begin{equation} 
\label{eq:deformation}
\widehat{\mathcal{V}}|_{\text{quad.}} = \sum_{\alpha\;\in\;\text{multiplets}}\,\bigl(\widehat{Q}\mathbb{X}^\alpha_0 \;\; \mathbb{X}^\alpha_1\bigr) \; \begin{pmatrix} D_{00} & D_{01} \\ D_{10} & D_{11} \end{pmatrix}
\begin{pmatrix} \mathbb{X}^\alpha_0  \\ \widehat{Q}\mathbb{X}^\alpha_1 \end{pmatrix} \, , 
\end{equation}
for a special basis of bosonic and fermionic fields $\{ \mathbb{X}^\alpha_0, \mathbb{X}^\alpha_1 \}$ and their $\widehat{Q}$-images. This split of fields is sometimes called the cohomological split, and there is an algorithmic procedure for determining the sets $\{ \mathbb{X}^\alpha_0, \mathbb{X}^\alpha_1 \}$ described in \cite{Jeon:2018kec}. Then, by acting with $\widehat{Q}$ on $\widehat{\mathcal{V}}$, one finds that the one-loop determinant for the operator $\widehat{Q} \widehat{\mathcal{V}}$ is given by
\begin{equation} 
\label{eq:Z1loop}
Z_\text{1-loop} = \sqrt{\frac{\det_\text{Coker$D_{10}$}(\widehat{Q}{}^2)}{\det_\text{Ker$D_{10}$}(\widehat{Q}{}^2)}} \, , 
\end{equation}
assuming we are dealing with real fields, or the square of the right hand side in the complex case. Therefore, for a given eigenvalue of $H := \widehat{Q}{}^2$, we need to know the dimensions of the kernel and cokernel of the $D_{10}$ operator.  The contribution of the other operators in \eqref{eq:deformation} to $Z_\text{1-loop}$ will drop out, and the choice of basis $\{ \mathbb{X}^\alpha_0, \mathbb{X}^\alpha_1 \}$ makes this cancellation explicit. The difference of dimensions just mentioned is encoded in the equivariant index of the $D_{10}$ operator,
\begin{equation} 
\label{eq:indD10}
\text{ind}_H(D_{10})(t) := \text{Tr}_\text{Ker$D_{10}$}  \, e^{\mathrm{i}Ht} - \text{Tr}_\text{Coker$D_{10}$}  \, e^{\mathrm{i}Ht} = \sum_{\mathfrak{n}} (m_{\mathfrak{n}}^{(0)} - m_{\mathfrak{n}}^{(1)})\,e^{\lambda_{\mathfrak{n}}t} \, . 
\end{equation}
Above, $m_{\mathfrak{n}}^{(0)}$ and $m_{\mathfrak{n}}^{(1)}$ are the dimensions of the kernel and cokernel of $D_{10}$ for a given eigenvalue $\lambda_{\mathfrak{n}}$ of the $\mathrm{i}H$ operator labeled by $\mathfrak{n}$, and $t$ is a formal expansion parameter. Knowing the equivariant index of $D_{10}$, the one-loop determinant is read off from \eqref{eq:Z1loop}:
\begin{equation}
\label{eq:1-loop-formal}
Z_\text{1-loop} = \prod_{\mathfrak{n}} \lambda_{\mathfrak{n}}{}^{\frac12(m_{\mathfrak{n}}^{(1)} - m_{\mathfrak{n}}^{(0)})} \, .
\end{equation}
Note that the infinite product above is \textit{a priori} only a formal expression, and we may need to introduce a suitable regulator. We will discuss this point later on.

Let us finish by pointing out that, apart from the explicit evaluation of the multiplicity of eigenvalues in the kernel and cokernel in the above formula, there are other ways of determining the equivariant index of the operator $D_{10}$ that we employ and discuss in detail in the coming sections. One way is via the general Atiyah-Singer index theorem for (elliptic) differential operators, while the other is via the Atiyah-Bott fixed point theorem that makes use of the fixed points of the action of $H$ (see \cite{Atiyah-book} for an introduction to the topic). In this context we recall that at a hands-on level, the so-called symbol of the differential operator $D_{10}$, denoted by $\sigma[\,D_{10}\,]$, is obtained by replacing derivatives with momenta $\partial_a \rightarrow p_a$. This notion is useful since, if two differential operators have the same symbol they have the same index\footnote{As stressed earlier, the mathematical theorems that we use here are valid for differential operators on compact spaces. We assume the same statements continue to hold in our non-compact case as long as we impose the relevant boundary conditions.}, which in practice means that one can often relate the problem of evaluating the index of $D_{10}$ to the evaluation of the index of some well-known differential operator.

\section{Vector multiplets}
\label{sec:vectors}

In this section we perform the steps required for the evaluation of the one-loop determinant for generic abelian vector multiplets, starting from the rewriting of the supersymmetry variations in terms of twisted variables, then performing the gauge-fixing, and proceeding with the evaluation of the $D_{10}$ operator and its index using three different methods. As in section \ref{subsec:review}, we refrain from reviewing the full 4d $\mathcal{N}=2$ Euclidean conformal supergravity formalism used in this paper, instead referring the reader to \cite{deWit:2017cle} for a thorough presentation. We merely gather some relevant aspects of the near-horizon background in appendix \ref{app:A} (see \cite{Hristov:2018lod} for more details). Some of the technical calculations pertaining to this section have also been relegated to appendix \ref{app:B}.

\subsection{Susy transformations and twisted variables}

We consider $n_V + 1$ vector multiplets coupled to the conformal supergravity background. The transformation rules for a Euclidean vector multiplet $\mathbb{V}^I$ under the localizing supercharge $Q$ parameterized by commuting symplectic-Majorana Killing spinors $(\xi^i,\,\kappa^i)$ (see \eqref{eq:loc-CKS} for their explicit expressions) are given by \cite{deWit:2017cle}
\begin{align}
\begin{split}
\label{eq:vector-4D}
Q X^I_\pm=&\; \pm\,\bar\xi_{i\pm}\,\Omega^{i\,I}_\pm \, , \\[1mm]
Q W^I_\mu=&\; \mathrm{i}\,\bar\xi_{i-}\,\gamma_\mu\,\Omega^{i\,I}_+ - \mathrm{i}\,\bar\xi_{i+}\,\gamma_\mu\,\Omega^{i\,I}_- \, , \\[1mm]
Q \Omega^{i\,I}_\pm =&\; -2\mathrm{i}\,\slash{\partial} X^I_\pm\,\xi^i{\!}_\mp - \tfrac12\bigl[F(W)_{ab}^{\mp\,I} - \tfrac14 X^I_\mp\,T_{ab}^\mp\,\bigr]\gamma^{ab} \xi^i{\!}_\pm - \varepsilon_{kj}\, Y^{ik\,I}\xi^j{\!}_\pm + 2\,X^I_\pm\,\kappa^i{\!}_\pm \, , \\[1mm] 
Q Y^{ij\,I} =&\; 2\mathrm{i}\,\varepsilon^{k(i}\,\bar{\xi}_{k-}\Slash{\mathcal{D}}\Omega^{j)\,I}_+ - 2\mathrm{i}\,\varepsilon^{k(i}\,\bar{\xi}_{k+}\Slash{\mathcal{D}}\Omega^{j)\,I}_- \, ,
\end{split}
\end{align}
with $I=0\ldots n_V$, the subscripts on spinors denote chiral projections and the bar is the standard Hermitian conjugate. The scalar fields $X_+^I$ and $X_-^I$ are real independent fields. Note that the action of $Q$ as defined above is (pseudo-)real, as can be checked using the standard rules for commuting symplectic-Majorana spinors laid out in \cite{Hristov:2018lod}. Owing to the conformal supergravity formalism, the algebra of $Q$ closes off-shell according to \eqref{eq:Q-alg}.

It will be convenient to change variables in each vector multiplet $\mathbb{V}^I$ and work with the so-called twisted gaugini defined by
\begin{equation}
\label{eq:twisted-fermions}
\lambda^I := \bar{\xi}_i\,\gamma^5\,\Omega^{i\,I} \, , \quad \lambda^I_\mu := \mathrm{i}\,\bar{\xi}_i\,\gamma_\mu\gamma^5\,\Omega^{i\,I} \, , \quad \lambda^{ij\,I} := \varepsilon^{k(i}\,\bar{\xi}_k\,\Omega^{j)\,I} \, .
\end{equation}
Note that these bilinears are (pseudo-)real. The relations above can be inverted by means of the Fierz identity:
\begin{equation}
\label{eq:twist-inverse}
\Omega^{i\,I}_\pm = \pm\,K^{-1}\bigl(\xi^i_\pm\,\lambda^I - \mathrm{i}\,\gamma^\mu\xi^i_\mp\,\lambda^I_\mu \mp 2\,\varepsilon_{jk}\,\xi^k_\pm\,\lambda^{ij\,I}\bigr) \, ,
\end{equation}
the prefactor being the norm of the Killing spinor
\begin{equation}
\label{eq:K-norm}
K := \bar{\xi}_k\,\xi^k = 2\cosh\eta \, .
\end{equation}
In particular, $K$ is nowhere vanishing so the change of variables to the twisted gaugini is regular. This bilinear will play a central role in the following computations.

The $Q$-supersymmetry transformations of the twisted gaugini are obtained from \eqref{eq:vector-4D},
\begin{align}
\begin{split}
Q\lambda^I =&\; \mathcal{L}_{\mathring{v}}X^I_+ + \mathcal{L}_{\mathring{v}}X^I_- \, , \\[1mm]
Q\lambda^I_\mu =&\; \mathcal{L}_{\mathring{v}}W^I_\mu + \partial_\mu\Lambda^I \, , \\[1mm]
Q\lambda^{ij\,I} =&\; \tfrac12\,K\,Y^{ij\,I} - \tfrac12\bigl(\varepsilon^{k(i}\bar{\xi}_k\gamma^{\mu\nu}\xi^{j)}\bigr)F^I_{\mu\nu} \\ 
&+ \tfrac{\mathrm{i}}{\sqrt{v_1}}\,\sigma_3{}^{(i}{}_k\varepsilon^{j)k}\,\bigl(X^I_+ - X^I_-\bigr) - 2\,\bigl(\mathrm{i}\,\varepsilon^{k(i}\bar{\xi}_{k+}\gamma^\mu\xi^{j)}_-\bigr)\,\partial_\mu\bigl(X^I_+ + X^I_-\bigr) \, ,
\end{split}
\end{align}
where we made use of the explicit values of the background $T$-tensor and the spinor $\kappa^i$ in \eqref{eq:1/2-BPS-T-tensor} and \eqref{eq:loc-CKS}. We will denote the spinor bilinears appearing on the right-hand side of the $\lambda^{ij\,I}$ variation by
\begin{equation}
\label{eq:K-bilin}
K_{\mu\nu}^{ij} := \varepsilon^{k(i}\bar{\xi}_k\gamma_{\mu\nu}\xi^{j)} \, , \qquad K_\mu^{ij} := \mathrm{i}\,\varepsilon^{k(i}\bar{\xi}_{k+}\gamma_\mu\xi^{j)}_- \, .
\end{equation}
These bilinears are pseudo-real, $\bigl(K_{\mu\nu}^{ij}\bigr)^\dagger = \varepsilon_{ik}\varepsilon_{jl}\,K_{\mu\nu}^{kl}\,$ and $\,\bigl(K_\mu^{ij}\bigr)^\dagger = \varepsilon_{ik}\varepsilon_{jl}\,K_\mu^{kl}$.

\subsection{Gauge fixing and ghosts}

We now proceed to fix the $U(1)$ gauge symmetry in each vector multiplet $\mathbb{V}^I$ by introducing the appropriate ghost fields, which we gather in a BRST complex. To do so, we introduce a set of ghosts fields $c^I$, anti-ghosts fields $b^I$ and Lagrange multiplier fields $B^I$. Then we introduce a standard BRST operator $Q_B$ acting on the fields as follows:
\begin{align}
\begin{split}
&\;Q_B W_\mu^I = \partial_\mu c^I \, , \;\; Q_B X_\pm^I = Q_B \Omega^{I\,i}_\pm = Q_B Y^I_{ij} = 0 \, , \\[1mm]
&\;Q_B b^I = B^I \, , \quad\;\;\; Q_B c^I = 0 \, , \quad\;\;\; Q_B B^I = 0 \, .
\end{split}
\end{align}
The gauge fields $W_\mu^I$ are the only vector multiplet fields transforming under $Q_B$ since the other fields are in the adjoint representation of the gauge group and we have an abelian symmetry.
With these transformation rules, it is straightforward to check that the algebra of the BRST supercharge is the standard nilpotent algebra $Q_B{}^2 = 0$.

We should also give appropriate $Q$-transformations to the ghost system. Following a standard procedure, we take
\begin{equation}
\label{eq:susy-ghosts}
Qc^I = -\Lambda^I \, , \quad Qb^I = 0 \, , \quad QB^I = \mathcal{L}_{\mathring{v}} b^I \, .
\end{equation}
These transformations are chosen so that the combined supercharge 
\begin{equation}
\widehat{Q} := Q + Q_B \, , 
\end{equation}
satisfies the algebra
\begin{equation}
\label{eq:gf-alg}
\widehat{Q}{}^2 = \mathcal{L}_{\mathring{v}} + \delta_{\mathrm{R}}(\Lambda^i{}_j) \, ,
\end{equation}
with parameters given in \eqref{eq:KV} and \eqref{eq:SU2-param}, as can be checked by an explicit calculation. In particular, the transformations \eqref{eq:susy-ghosts} ensure that the field-dependent gauge transformation in \eqref{eq:Q-alg} is canceled by the cross term $Q\,Q_B$ present in $\widehat{Q}{}^2$.

To summarize, the transformation rules for the fields of $\mathbb{V}^I$ and the ghost fields under the supercharge $\widehat{Q}$ can be written in terms of twisted fermions $\lambda^I$, $\lambda^I_\mu$ and $\lambda^{ij\,I}$ as follows:
\begin{align}
\begin{split}
\label{eq:Qhat-transfo}
&\!\!\!\widehat{Q}W^I_\mu = \lambda^I_\mu + \partial_\mu c^I \, , \qquad\; \widehat{Q}X^I_\pm = K^{-1}\,\bigl(\pm \tfrac12\,\mathring{v}^\mu\lambda^I_\mu +  K_\pm\,\lambda^I\bigr) \, , \\[1mm]
&\!\!\!\widehat{Q}\lambda^I_\mu = \mathcal{L}_{\mathring{v}}W^I_\mu + \partial_\mu\Lambda^I \, , \quad \widehat{Q}\lambda^I = \mathcal{L}_{\mathring{v}}\bigl(X^I_+ - X^I_-\bigr)\, ,  \\[1mm]
&\!\!\!\widehat{Q}\lambda^{ij\,I} = \tfrac12\,K\,Y^{ij\,I} - \tfrac12\,K_{\mu\nu}^{ij}\,F^{\mu\nu\,I} + \tfrac{\mathrm{i}}{\sqrt{v_1}}\,\sigma_3{}^{(i}{}_k\varepsilon^{j)k}\,\bigl(X^I_+ - X^I_-\bigr) - 2\,K_\mu^{ij}\,\partial^\mu\bigl(X^I_+ + X^I_-\bigr) \, , \\[1mm]
&\!\!\!\widehat{Q}c^I = -\Lambda^I \, ,\quad \widehat{Q}b^I = B^I \, , \quad \widehat{Q}B^I = \mathcal{L}_{\mathring{v}}b^I \, .
\end{split}
\end{align}
Here we have introduced the chiral projections of $K$ defined in \eqref{eq:K-norm}, $K_\pm := \bar{\xi}_{k\pm} \xi^k{\!}_\pm$. They already appeared in \eqref{eq:Lambda-param} and are related to $K$ as $K = K_+ + K_-$. The auxiliary fields $Y^{ij\,I}$ also transform under $\widehat{Q}$, although the explicit form of their transformation will not be needed in what follows.

\subsection{Reality conditions and the $D_{10}$ operator}
Examining \eqref{eq:Qhat-transfo}, we see that the scalar fields naturally appear in combinations $X^I_+ \pm X^I_-$. We will accordingly write the $\widehat{Q}$-transformation rules in terms of
\begin{equation}
\sigma^I := -\tfrac12\mathrm{i}\,(X^I_+ + X^I_-) \, , \qquad \rho^I := \tfrac12\,(X^I_+ - X^I_-) \, , \qquad \tilde{W}^I_\mu := -\mathrm{i}\,W^I_\mu \, ,
\end{equation}
with prefactors chosen for later convenience. Importantly, we will use the following reality conditions for the fields $(\sigma^I,\,\rho^I,\,\tilde{W}_\mu^I)$ when computing the one-loop determinant: 
\begin{equation}
\label{eq:rot-real-cond}
\sigma^I{\,}^\dagger = \sigma^I \, ,\qquad \rho^I{\,}^\dagger = \rho^I \, , \qquad \tilde{W}^I_\mu{\,}^\dagger = \tilde{W}^I_\mu \, .
\end{equation}
This choice of contour corresponds to a rotation of the original fields $X^I_+ + X^I_-$ and $W^I_\mu$ in field space, while keeping $X_+^I - X_-^I$ real. It follows from the choice of contour already used in \cite{Hristov:2018lod} to obtain the localization locus and evaluate the contribution of the classical action to the localized path-integral \eqref{eq:first-macroscopic-degeneracy}. In this basis, \eqref{eq:Qhat-transfo} reads
\begin{align}
\begin{split}
\label{eq:Qhat-transfo-good-basis}
&\;\widehat{Q}\tilde{W}^I_\mu = -\mathrm{i}\,\lambda^I_\mu - \mathrm{i}\,\partial_\mu c^I \, , \qquad \widehat{Q}\lambda_\mu^I = \mathrm{i}\,\mathring{v}^\nu\partial_\nu \tilde{W}^I_\mu + \partial_\mu\bigl(2\,K\,\rho^I + 4\mathrm{i}\,\sigma^I - \mathrm{i}\,\mathring{v}^\nu \tilde{W}^I_\nu\bigr) \, , \\[1mm]
&\;\widehat{Q}\sigma^I = -\tfrac12\mathrm{i}\,\lambda^I \, , \qquad \widehat{Q}\rho^I = K^{-1}\bigl(\tfrac12\,\mathring{v}^\mu\lambda^I_\mu - \lambda^I\bigr) \, , \qquad \widehat{Q}\lambda^I = 2\mathrm{i}\,\mathring{v}^\mu\partial_\mu\sigma^I \, , \\[1mm]
&\;\widehat{Q}\lambda^{ij\,I} = \tfrac12\,K\,Y^{ij\,I} - \tfrac12\mathrm{i}\,K_{\mu\nu}^{ij}\,\tilde{F}^{\mu\nu\,I} + \tfrac{2}{\sqrt{v_1}}\mathrm{i}\,\sigma_3{}^{(i}{}_k\varepsilon^{j)k}\,\rho^I - 4\mathrm{i}\,K_\mu^{ij}\,\partial^\mu\sigma^I \, , \\[1mm]
&\;\widehat{Q}c^I = -(2\,K\,\rho^I + 4\mathrm{i}\,\sigma^I - \mathrm{i}\,\mathring{v}^\mu \tilde{W}^I_\mu) \, ,\qquad \widehat{Q}b^I = B^I \, , \qquad \widehat{Q}B^I = \mathring{v}^\mu\partial_\mu b^I \, ,
\end{split}
\end{align}
where we used the explicit expression of the parameter of the gauge transformation $\Lambda^I$ in terms of the fields of $\mathbb{V}^I$ given in \eqref{eq:Lambda-param}. Of course, due to our choice of rotated reality conditions \eqref{eq:rot-real-cond}, the action of $\widehat{Q}$ is no longer (pseudo-)real. 

We now further split the fields into the following sets \cite{Jeon:2018kec}:
\begin{equation}
\label{eq:vec-split}
\mathbb{X}^I_0 := \{\sigma^I\,,\tilde{W}^I_\mu\} \, , \quad \mathbb{X}^I_1 := \{\lambda^{ij\,I}\,,c^I\,,b^I\} \, ,
\end{equation}
and their $\widehat{Q}$-images $\widehat{Q}\mathbb{X}^I_0$, $\widehat{Q}\mathbb{X}^I_1$. This so-called cohomological split is particularly useful for the computation of one-loop determinants since it allows us to isolate the differential operator $D_{10}$, as already explained around \eqref{eq:deformation}. 

To identify the operator $D^\mathrm{vec}_{10}$ relevant to the vector multiplets, we go back to the fermionic deformation $\mathcal{V}^\mathrm{vec}$ used for localization in \cite{Hristov:2018lod}. There it was written in terms of the gaugini $\Omega^{i\,I}$, and we should now also include the relevant ghost terms to fix the gauge as first explained in \cite{Pestun:2007rz}. Doing so we obtain the following fermionic deformation:
\begin{equation}
\widehat{\mathcal{V}}^\mathrm{vec} = \int d^4x \, \frac{\sqrt{\mathring{g}}}{K} \, \sum_I \, \Bigl[\Omega^{i\,I}_+\bigl(\widehat{Q}\Omega^{i\,I}_+\bigr)^\dagger + \Omega^{i\,I}_-\bigl(\widehat{Q}\Omega^{i\,I}_-\bigr)^\dagger + K\,b^I G(W^I) \Bigr] \, ,
\end{equation}
in which we use the Hermitian conjugate as defined in \eqref{eq:rot-real-cond} to build the inner product, and we leave the gauge-fixing function $G(W^I)$ unspecified for now. Note that we have included an extra factor of $K^{-1}$ compared to \cite{Hristov:2018lod}. This factor is nowhere vanishing and hence does not modify the analysis of the localization locus performed in \cite{Hristov:2018lod}. It will however allow us to use integration by parts when discussing the $D^\mathrm{vec}_{10}$ operator contained in $\widehat{\mathcal{V}}^{\rm vec}|_{\text{quad.}}$, as will be discussed in due course. In terms of the twisted fermions \eqref{eq:twisted-fermions},
\begin{equation}
\label{eq:Vhat}
\widehat{\mathcal{V}}^\mathrm{vec} = \int d^4x \, \frac{\sqrt{\mathring{g}}}{K^2}\,\sum_I\,\Bigl[\lambda^I(\widehat{Q}\lambda^I)^\dagger + \lambda^{I\,\mu}(\widehat{Q}\lambda^I_\mu)^\dagger + 2\,\lambda^{I\,ij}(\widehat{Q}\lambda^{I\,ij})^\dagger +  K^2\,b^I G(W^I)\Bigr] \, .
\end{equation}

Using \eqref{eq:Qhat-transfo-good-basis}, we can compute the terms relevant to $D^\mathrm{vec}_{10}$ in a given multiplet $\mathbb{V}^I$ as follows. The first term in the deformation can be written in terms of the fields in $\mathbb{X}^I_0$, $\mathbb{X}^I_1$ and their $\widehat{Q}$-images as
\begin{equation}
\lambda^I\,(\widehat{Q}\lambda^I)^\dagger = (2\mathrm{i}\,\widehat{Q}\sigma^I)(-2\mathrm{i}\,\mathring{v}^\mu\partial_\mu\sigma^I) \, .
\end{equation}
The right-hand side involves fields in $\widehat{Q}\mathbb{X}^I_0$ and $\mathbb{X}^I_0$, and therefore this term contributes to the $D^\mathrm{vec}_{00}$ operator but not to $D^\mathrm{vec}_{10}$. To obtain the contribution of the term $\lambda^{\mu\,I}(\widehat{Q}\lambda^I_\mu)^\dagger$ in \eqref{eq:Vhat}, we note that with the choice of reality conditions \eqref{eq:rot-real-cond},
\begin{equation}
(\widehat{Q}\lambda^I_\mu)^\dagger = -\mathrm{i}\,\mathring{v}^\nu\partial_\nu \tilde{W}^I_\mu - \partial_\mu\bigl(\widehat{Q}c^I + 8\mathrm{i}\,\sigma^I - 2\mathrm{i}\,\mathring{v}^\nu \tilde{W}^I_\nu\bigr) \, .
\end{equation}
Since $\lambda^{\mu\,I} = \mathrm{i}\,\widehat{Q}\tilde{W}^{\mu\,I} - \partial^\mu c^I$, the second term in the deformation contributes
\begin{equation}
\lambda^{\mu\,I}(\widehat{Q}\lambda^I_\mu)^\dagger  \; \ni \; (\partial^\mu c^I)\,\bigl(\mathrm{i}\,\mathring{v}^\nu\partial_\nu \tilde{W}^I_\mu - 2\mathrm{i}\,\partial_\mu(\mathring{v}^\nu \tilde{W}^I_\nu)  + 8\mathrm{i}\,\partial_\mu\sigma^I\bigr) \, ,
\end{equation}
to the $D^\mathrm{vec}_{10}$ operator. For the third term $2\,\lambda^{ij\,I}(\widehat{Q}\lambda^{ij\,I})^\dagger$, \eqref{eq:rot-real-cond} implies
\begin{equation}
(\widehat{Q}\lambda^{ij\,I})^\dagger = \varepsilon_{ik}\varepsilon_{jl}\,\bigl(\widehat{Q}\lambda^{kl\,I} + \mathrm{i}\,K_{\mu\nu}^{kl}\,\tilde{F}^{\mu\nu\,I} + 8\mathrm{i}\,K_\mu^{kl}\,\partial^\mu\sigma^I\bigr) \, ,
\end{equation}
which yields a contribution to the $D^\mathrm{vec}_{10}$ operator of
\begin{equation}
2\,\lambda^{ij\,I}(\widehat{Q}\lambda^{ij\,I})^\dagger \; \ni \; 2\,\lambda^{ij\,I}\bigl(\mathrm{i}\,K^{\mu\nu}_{ij}\,\tilde{F}^I_{\mu\nu} + 8\mathrm{i}\,K^\mu_{ij}\,\partial_\mu\sigma^I\bigr) \, .
\end{equation}
Putting the above contributions together, the operator $D^\mathrm{vec}_{10}$ is given explicitly by
\begin{align}
\label{eq:D10}
&\mathbb{X}^I_1\,D^\mathrm{vec}_{10}\,\mathbb{X}^I_0 = \\
&\;\frac{\sqrt{\mathring{g}}}{K^2}\Bigl[(\partial^\mu c)\,\bigl(\mathring{v}^\nu\partial_\nu \tilde{W}_\mu - 2\,\partial_\mu(\mathring{v}^\nu \tilde{W}_\nu)  + 8\,\partial_\mu\sigma\bigr) + 2\,\lambda^{ij}\bigl(K^{\mu\nu}_{ij}\,\tilde{F}_{\mu\nu} + 8\,K^\mu_{ij}\,\partial_\mu\sigma\bigr) + K^2 b\,G(\tilde{W})\Bigr] \nonumber \, ,
\end{align}
where we have dropped an irrelevant factor of $\mathrm{i}$, and refrained from writing the vector multiplet index $I$ on the fields on the right-hand-side to lighten the notation.

Having identified $D^\mathrm{vec}_{10}$, we proceed with the computation of its equivariant index. We will do so using three different methods. First, following \cite{David:2016onq}, we will compute the dimensions of its kernel and cokernel directly. This requires a precise discussion of boundary conditions used in the analysis of the kernel and cokernel of the differential operator \eqref{eq:D10}, and will yield concrete expressions for the modes giving a non-trivial contribution to the vector multiplet one-loop determinant $Z_\text{1-loop}^\mathrm{vec}$. Afterwards we will use the Atiyah-Singer theorem to compute the equivariant index of $D^\mathrm{vec}_{10}$ in terms of topological quantities. Lastly, we will also make use of the Atiyah-Bott fixed point theorem after suitably deforming the $\widehat{Q}^2$-action to include refinement. As we will see all three methods yield the same result, hence giving a consistency check of the computation as well as allowing us to discuss various generalizations and relations with previous results in the literature.

\subsection{Method I: Mode analysis}
\label{sec:modes}

Before deriving and analyzing the kernel and cokernel equations for the differential operator $D^\mathrm{vec}_{10}$, we discuss the boundary conditions of the various fields that will play a role. This is important in what follows, as we are going to look for solutions to the equations in the specific field subspace specified by these boundary conditions.

\subsubsection{Mode expansion and boundary conditions}
\label{sec:mod-exp-bc}

To establish a set of admissible boundary conditions, we proceed along the lines of \cite{David:2016onq}. The so-called \emph{normalizable boundary conditions} stem from requiring that the Gaussian path-integral is normalized,
\begin{equation}
\label{eq:norm-Gauss}
\int \mathcal{D}\Phi\,\exp\Bigl[\,-\int \mathrm{d}^4 x\,\sqrt{\mathring{g}}\;|\Phi|^2\,\Bigr] = 1 \, ,
\end{equation}
where $\Phi$ denotes any field. To analyze the kernel and cokernel equations, we will decompose the fields in $\mathbb{X}_0^I$ and $\mathbb{X}_1^I$ in Fourier modes along the $\mathbb{H}_2$ and $S^2$ factors of the near-horizon geometry \eqref{eq:NHG-metric}. For a generic scalar field $S$ (from the point of view of the 2-sphere), we can use the standard spherical harmonics $Y_\ell{}^m$ to expand in radial modes and along the Euclidean time circle,
\begin{equation}
\label{eq:scal-mod-exp}
S = S_{(n-s/2,\,\ell,\,m)}(\eta)\,e^{\mathrm{i}(n - s/2)\tau}\,Y_\ell{}^m(\theta,\varphi) \, .
\end{equation}
Here $s$ is the charge under the R-transformation in the algebra of $\widehat{Q}^2$ \eqref{eq:gf-alg} (for the explicit expressions pertaining to vector multiplet fields, see appendix \ref{app:B}). We introduce this quantum number directly in the mode decomposition along the Euclidean time circle so that the action of $\widehat{Q}^2$ on a generic field $\Phi$ takes a universal form regardless of the R-charge of the field,\footnote{The reason why it is possible to combine the Lie derivative and the R-transformation is because the SU(2)$_{\rm R}$ symmetry is broken to $U(1)_{\rm R}$ on the half-BPS background.}
\begin{equation}
\label{eq:H-eigen}
\widehat{Q}{}^2 \Phi = \frac{2\mathrm{i}n}{\sqrt{v_1}}\,\Phi \, .
\end{equation}
For a vector $V_\alpha$ with $x^\alpha = (\theta,\varphi)$, we expand along vector spherical harmonics using the basis put forward in \cite{Banerjee:2011jp}: given a set of normalized eigenfunctions $\{U_k\}$ of the scalar Laplacian on the 2-sphere $(-\nabla_{S^2}^2)$ with eigenvalues $\kappa^{(k)}$, a normalized basis for vector fields on $S^2$ is given by
\begin{equation}
\frac{1}{\sqrt{\kappa^{(k)}}}\,\partial_\alpha U_k \, , \qquad \textnormal{and} \qquad \frac{1}{\sqrt{\kappa^{(k)}}}\,\varepsilon_{\alpha\beta}\,\partial^\beta U_k \, , 
\end{equation}
with the invariant anti-symmetric tensor  on $S^2$ given by $\varepsilon_{\theta\varphi} = v_2\,\sin\theta$ in the coordinates \eqref{eq:NHG-metric}. Since the spherical harmonics $Y_\ell{}^m$ are eigenfunctions of the scalar Laplacian, we may expand $V_\alpha$ as
\begin{equation}
\label{eq:vec-mod-exp}
V_\alpha = V_B(\eta)\,e^{\mathrm{i}(n - s/2)\tau}\,\partial_\alpha Y_\ell{}^m(\theta,\varphi) + V_C(\eta)\,e^{\mathrm{i}(n - s/2)\tau}\,\varepsilon_{\alpha\beta}\,\partial^\beta Y_\ell{}^m(\theta,\varphi) \, ,
\end{equation}
where we have absorbed the normalization of $\sqrt{\kappa^{(\ell,m)}}$ in the functions $V_B(\eta)$ and $V_C(\eta)$, and here and below we omit the quantum number labels on the radial modes for convenience of notation.

In the above mode decomposition, the condition \eqref{eq:norm-Gauss} amounts to requiring that the radial modes are such that $\sinh\eta\,|S(\eta)|^2$, $\sinh\eta\,|V_B(\eta)|^2$, $\sinh\eta\,|V_C(\eta)|^2$ decay fast enough when $\eta \rightarrow \infty$. Thus, the normalizable boundary conditions for the bosonic fields are
\begin{equation}
\label{eq:norm-bc-vec-bos}
e^{\eta/2}\sigma(\eta) \rightarrow 0 \, , \quad e^{-\eta/2}\tilde{W}_\tau(\eta)  \rightarrow 0 \, , \quad e^{\eta/2}\tilde{W}_{\hat{\mu}}(\eta) \rightarrow 0 \, ,
\end{equation}
when $\eta \rightarrow \infty$, where $x^{\hat{\mu}} = (\eta,\theta,\varphi)$. Having established the boundary conditions on the bosonic fields, we impose conditions on the fermions which are consistent with supersymmetry. Using the transformation rules \eqref{eq:Qhat-transfo-good-basis} and the behavior \eqref{eq:norm-bc-vec-bos}, we see that $e^{-\eta/2}\widehat{Q}c \rightarrow 0$ so we should require $e^{-\eta/2}c(\eta) \rightarrow 0$. However, since the ghost field acts as a gauge transformation parameter for the gauge field, we must also require that it does not change the asymptotic behavior of $\tilde{W}_\mu$. This leads to a stronger condition at infinity \cite{David:2016onq},
\begin{equation}
\label{eq:susy-bc-ghost}
c(\eta) \sim \mathcal{O}(1) + o(e^{-\eta/2}) \, , \quad b(\eta) \sim \mathcal{O}(1) + o(e^{-\eta/2}) \, ,
\end{equation}
where we impose the same asymptotic behavior on the anti-ghost since $b$ and $c$ are paired in the ghost Lagrangian. For the twisted fermions $\lambda^{ij}$, the transformations \eqref{eq:Qhat-transfo-good-basis} together with \eqref{eq:norm-bc-vec-bos} show that $e^{-\eta/2}\widehat{Q}\lambda^{ij} \rightarrow 0$ when $\eta \rightarrow \infty$. Thus, we impose
\begin{equation}
\label{eq:susy-bc-vec-ferm}
e^{-\eta/2}\lambda^{ij}(\eta) \rightarrow 0 \, .
\end{equation}
The asymptotic behavior of $\lambda(\eta)$ and $\lambda_\mu(\eta)$ can be obtained in a similar fashion, although we will not need them in what follows. As a remark, note that the above supersymmetric boundary conditions are weaker than the normalizable boundary conditions for the fermions. Indeed, with the normalizable condition $e^{\eta/2}\Omega^i(\eta) \rightarrow 0$ from \eqref{eq:norm-Gauss}, using the expression for the twisted fermions in terms of the gaugini \eqref{eq:twist-inverse} and the explicit form of the $\xi^i$ spinor in appendix \ref{app:A}, we obtain $\lambda^{ij}(\eta) \rightarrow 0$. This is stronger than, and therefore implies, \eqref{eq:susy-bc-vec-ferm}.

We also impose smoothness conditions near the origin $\eta \rightarrow 0$, which follow from requiring that the Wilson line in the definition of the QEF \eqref{eq:first-macroscopic-degeneracy} be contractible at the origin $\eta = 0$. The smooth behavior near the origin will depend on the quantum number $n$ for the various radial modes. For the fields in $\mathbb{X}_0^I$ we require \cite{David:2016onq}
\begin{align}
\label{eq:smooth-vec-bos}
\tilde{W}^{(n\neq0,\ell)}_\tau(\eta) \sim \eta^{|n|} \, ,  \quad &\tilde{W}^{(n=0,\ell)}_\tau(\eta) \sim \eta^2 \, , \quad \tilde{W}^{(n\neq0,\ell)}_\eta(\eta) \sim \eta^{|n|-1} \, , \quad \tilde{W}^{(n=0,\ell)}_\eta(\eta) \sim \eta \, , \nonumber \\[1mm]
&\;\;\; \tilde{W}_\alpha(\eta) \sim \eta^{|n|} \, , \quad \sigma(\eta) \sim \eta^{|n|} \, ,
\end{align}
when $\eta \rightarrow 0$. Similarly, for the fields in $\mathbb{X}_1^I$,
\begin{equation}
\label{eq:smooth-vec-ferm}
c(\eta) \sim \eta^{|n|} \, , \quad b(\eta) \sim \eta^{|n|} \, , \quad \lambda^{ij}(\eta) \sim \eta^{|n|} \, .
\end{equation}
With these boundary and smoothness conditions, we proceed to analyze the equations giving the kernel and cokernel of the $D_{10}^\mathrm{vec}$ differential operator \eqref{eq:D10}.

\subsubsection{Kernel analysis}
\label{sec:kerneq}

To obtain the kernel equations, we first make a change of variables in the $\mathbb{X}^I_1$ field set\footnote{This will also appear in the index theorem computation, see \eqref{eq:bsW-shift} below.},
\begin{equation}
\label{eq:hatb}
\hat{b} := b + \mathring{v}^\mu\partial_\mu c \, .
\end{equation}
Note that the boundary conditions on $\hat{b}$ are the same as on the original field $b$ and given in \eqref{eq:susy-bc-ghost}. Varying \eqref{eq:D10} with respect to $c$, $\lambda^{ij}$, $\hat{b}$ and setting the result to zero yields the kernel equations. The kernel equation associated to the ghost field $c$ is
\begin{equation}
\frac{\delta}{\delta c} \;\; : \;\; \mathring{v}^\nu\partial_\mu\Bigl(\frac{\sqrt{\mathring{g}}}{K^2}\,\partial_\nu \tilde{W}^{\mu}\Bigr) - 2\,\partial_\mu\Bigl(\frac{\sqrt{\mathring{g}}}{K^2}\,\partial^\mu\bigl(\mathring{v}^\nu\tilde{W}_\nu - 4\,\sigma\bigr)\Bigr) - \mathring{v}^\mu\partial_\mu\Bigl(\sqrt{\mathring{g}}\,G(\tilde{W})\Bigr) = 0 \, ,
\end{equation}
where the last term comes from the change of variable \eqref{eq:hatb}. In deriving the above, we have used integration by parts. Due to our inclusion of a factor of $K^{-1}$ in $\hat{\mathcal{V}}^{\,\mathrm{vec}}$ \eqref{eq:Vhat} and the boundary conditions discussed above, the boundary terms vanish. We now observe that a convenient choice of gauge-fixing for the abelian gauge symmetry is
\begin{equation}
\label{eq:g-f}
G(\tilde{W}) = \nabla_\mu\Bigl(\frac{1}{K^2}\,\tilde{W}^\mu\Bigr) \, .
\end{equation}
In this gauge, the kernel equation associated to $c$ reduces to
\begin{equation}
\label{eq:kerneq-c}
\frac{\delta}{\delta c} \;\; : \quad \nabla_\mu\Bigl[\frac{1}{K^2}\,\partial^\mu\bigl(\mathring{v}^\nu \tilde{W}_\nu - 4\,\sigma\bigr)\Bigr] = 0 \, .
\end{equation}
Varying \eqref{eq:D10} with respect to $\lambda^{ij}$, we obtain the kernel equations
\begin{equation}
\frac{\delta}{\delta \lambda^{ij}} \;\; : \quad K^{\mu\nu}_{ij}\tilde{F}_{\mu\nu} + 8\,K^\mu_{ij}\,\partial_\mu\sigma = 0 \, .
\end{equation}
This can be rewritten slightly by means of the Fierz identity, which can be used to show that $4\,K^\mu_{ij} = K^{\mu\nu}_{ij}\mathring{v}_\nu$. Thus, we have
\begin{equation}
\label{eq:kerneq-lambda}
\frac{\delta}{\delta \lambda^{ij}} \;\; : \quad K^{\mu\nu}_{ij}\Bigl(\tilde{F}_{\mu\nu} -2\,\mathring{v}_\mu \partial_\nu\sigma\Bigr) = 0 \, .
\end{equation}
Lastly, the kernel equation associated to the field $\hat{b}$ is simply the gauge-fixing condition for the vector field $\tilde{W}_\mu$,
\begin{equation}
\label{eq:kerneq-b}
\frac{\delta}{\delta\hat{b}} \;\; : \quad \nabla_\mu\Bigl(\frac{1}{K^2}\,\tilde{W}^\mu\Bigr) = 0 \, .
\vspace{1.5mm}
\end{equation}

In the gauge \eqref{eq:g-f}, the solutions of \eqref{eq:kerneq-c}, \eqref{eq:kerneq-lambda} and \eqref{eq:kerneq-b} subject to the boundary and smoothness conditions \eqref{eq:norm-bc-vec-bos} and \eqref{eq:smooth-vec-bos} furnish the kernel of $D_{10}^\mathrm{vec}$. We discuss the details of these solutions in appendix \ref{app:kernel-modes}. After expanding all the $\mathbb{X}_0^I$ fields in modes, the problem reduces to a set of ordinary differential equations on the radial modes. We examine these ODEs in detail and come to the conclusion that there are no non-trivial solutions compatible with \eqref{eq:norm-bc-vec-bos} and \eqref{eq:smooth-vec-bos}. Thus, we conclude that in the subspace specified by the boundary and smoothness conditions, the kernel of $D_{10}^\mathrm{vec}$ is empty.

\subsubsection{Cokernel analysis}
\label{sec:cokerneq}

The cokernel equations are obtained by varying \eqref{eq:D10} with respect to $\tilde{W}_\mu$ and $\sigma$. We find
\begin{equation}
\label{eq:cokerneq-W}
\frac{\delta}{\delta \tilde{W}_\mu} \;\; : \quad \nabla_\nu\Bigl[\frac{1}{K^2}\,\Bigl(\mathring{v}^\mu\partial^\nu c + 2\,K_{ij}^{\mu\nu}\,\lambda^{ij}\Bigr)\Bigr] - \frac{1}{2K^2}\,\partial^\mu\hat{b} = 0 \, ,
\end{equation}
where we used the gauge-fixing function \eqref{eq:g-f}. Varying with respect to $\sigma$, we obtain
\begin{equation}
\label{eq:cokerneq-sigma}
\frac{\delta}{\delta \sigma} \;\; : \quad \nabla_\mu\Bigl[\frac{1}{K^2}\,\Bigl(2\,\partial^\mu c + K^{\mu\nu}_{ij}\,\mathring{v}_\nu\,\lambda^{ij}\Bigr)\Bigr] = 0 \, ,
\end{equation}
after using again the Fierz identity $4\,K^\mu_{ij} = K^{\mu\nu}_{ij}\mathring{v}_\nu$. Just as for the kernel analysis, the solutions of \eqref{eq:cokerneq-W} and \eqref{eq:cokerneq-sigma} subject to the boundary and smoothness conditions \eqref{eq:susy-bc-ghost}, \eqref{eq:susy-bc-vec-ferm} and \eqref{eq:smooth-vec-ferm} furnish the cokernel of $D_{10}^\mathrm{vec}$. The details are discussed in appendix \ref{app:cokernel-modes}, where we reduce the problem to a set of ODEs for the radial modes of the fermions. In contrast to the kernel case, we do find non-trivial solutions compatible with the boundary and smoothness conditions. An essential ingredient for this difference is that the ghost and antighost fields are allowed to go to a non-zero constant when $\eta \rightarrow \infty$. The number of solutions we find depends on the quantum numbers $(n,\ell)$ appearing in the mode decomposition, and the result for the real dimension of the cokernel of $D_{10}^\mathrm{vec}$ is
\begin{itemize}
\item $(n \neq 0,\,\ell \neq 0)$ : dim Coker\,$D_{10}^\mathrm{vec}$ = 0
\item $(n \neq 0,\,\ell = 0)$ : dim Coker\,$D_{10}^\mathrm{vec}$ = 1
\item $(n = 0,\,\ell = 0)$ : dim Coker\,$D_{10}^\mathrm{vec}$ = 2
\end{itemize}
We note that the same result for the kernel and cokernel was obtained in \cite{David:2016onq}, where the authors analyzed a three-dimensional situation analogous to ours with similar radial ODEs and identical boundary and smoothness conditions.

\subsubsection{Result}
\label{sec:mode-result}

Having obtained the dimensions of the kernel and cokernel of the $D_{10}^\mathrm{vec}$ operator, we use the general formalism reviewed in section \ref{subsec:looprev} to write the one-loop determinant for vector multiplets. According to \eqref{eq:H-eigen}, the eigenvalues of $\widehat{Q}^2$ are labeled by $\mathfrak{n} = \{n\} \in \mathbb{Z}$, 
\begin{equation}
\label{eq:H-eigen-2}
\widehat{Q}^2\,\mathbb{X}^I_{0,1} = \frac{2\mathrm{i}n}{\sqrt{v_1}}\,\mathbb{X}^I_{0,1} =: \lambda_n\,\mathbb{X}^I_{0,1} \, , 
\end{equation}
while the multiplicities $m_n^{(0)}$ and $m_n^{(1)}$ are given by
\begin{equation}
m_n^{(0)} = 0 \, , \qquad m_n^{(1)} = \begin{cases} 1 \;\; &\text{for} \;\; n \neq 0 \\ 2 \;\; &\text{for} \;\; n = 0 \end{cases} \; .
\end{equation}
It is clear from \eqref{eq:H-eigen-2} that the case $n=0$ corresponds to zero-modes of the Hamiltonian $\widehat{Q}^2$. This is also explained in appendix \ref{app:cokernel-modes}, where it is shown that the two solutions in the cokernel in the case $n=0$ are two constant modes for the ghost and anti-ghost fields, see \eqref{eq:cokernel-zm}. Constant modes are however not normalizable on the non-compact $\mathbb{H}_2 \times S^2$ space, and we therefore discard them from the determinant.\footnote{Another reason for not considering constant modes for the ghosts is that such modes are zero-modes of the $U(1)$ gauge transformation for the vector fields. In the Batalin-Vilkovisky quantization one then adds ghost-for-ghost fields to remove these zero-modes, see e.g. \cite{Murthy:2015yfa}.} Thus, from \eqref{eq:1-loop-formal} we obtain the one-loop determinant for a vector multiplet:
\begin{equation}
\label{eq:Z-modes}
Z^\mathrm{vec}_{\textnormal{1-loop}} = \prod_{n\in\mathbb{Z}^*}\,\Bigl(\frac{2\mathrm{i}n}{\sqrt{v_1}}\Bigr)^{1/2} = \prod_{n\geq1}\,\Bigl(\frac{4 n^2}{v_1}\Bigr)^{1/2} \, ,
\end{equation}
where we have taken into account all the modes with $n \in \mathbb{Z}^*$ in the decomposition along the Euclidean time circle. We will soon discuss a suitable regularization of the above infinite product. For the time being, we present another method to compute the equivariant index of $D_{10}^\mathrm{vec}$ based on the Atiyah-Singer theorem, which will lead to the same result \eqref{eq:Z-modes}.

\subsection{Method II: Atiyah-Singer index theorem}
\label{sec:index-result}

To apply the equivariant Atiyah-Singer index theorem, it will be convenient to relate the index of the operator $D^\mathrm{vec}_{10}$ defined in \eqref{eq:D10} to the index of some known differential operator.  Upon explicitly evaluating the bilinears $K^{\mu\nu}_{ij}$, we find that the symbol of $D^\mathrm{vec}_{10}$ (as discussed in section \ref{subsec:looprev}) is represented by the following matrix:
\begin{equation}
\label{eq:D10-symbol}
\mathbb{X}^I_1
\begin{pmatrix}
-8\,p^2 & 2\sinh\eta\,\bigl(p_1^2 + 2\,\vec{p}{\,}^2\bigr) & -2\sinh\eta\,p_1\,p_2 & -2\sinh\eta\,p_1\,p_3 & -2\sinh\eta\,p_1\,p_4 \\ 
0 & p_1 & p_2 & p_3 & p_4 \\
-2\sinh\eta\,p_2 & -p_2 & p_1 & -\cosh\eta\,p_4 & \cosh\eta\,p_3 \\
-2\sinh\eta\,p_3 & -p_3 & \cosh\eta\,p_4 & p_1 & -\cosh\eta\,p_2 \\
-2\sinh\eta\,p_4 & -p_4 & -\cosh\eta\,p_3 & \cosh\eta\,p_2 & p_1 
\end{pmatrix} 
\mathbb{X}^I_0\, ,
\end{equation}
with $\vec{p}{\,}^2=\sum_{i=2}^4 p_i^2$ and all indices are tangent space indices. Note that in order to write the symbol in the above form, we used the gauge-fixing function \eqref{eq:g-f} and reorganized the fermionic fields in the set $\mathbb{X}^I_1$,
\begin{equation}
\label{eq:symb-natural}
\mathbb{X}_0^I = \begin{pmatrix} \sigma^I \\ \tilde{W}^I_1 \\ \tilde{W}^I_2 \\ \tilde{W}^I_3 \\ \tilde{W}^I_4 \end{pmatrix} \, , \qquad 
\mathbb{X}_1^I = \begin{pmatrix} c^I \\ b^I \\ -8\mathrm{i}\,\lambda^{12\,I} \\ 4\mathrm{i}\bigl(e^{\mathrm{i}\tau}\lambda^{11\,I} - e^{-\mathrm{i}\tau}\lambda^{22\,I}\bigr) \\ 4\bigl(e^{\mathrm{i}\tau}\lambda^{11\,I} + e^{-\mathrm{i}\tau}\lambda^{22\,I}\bigr) \end{pmatrix}.
\end{equation}
These combinations of fields are precisely the ones appearing in the mode analysis of appendix \ref{app:cokernel-modes} in \eqref{eq:ferm_decomp} and \eqref{eq:mode-natural}. They are real and neutral under the R-transformation present in the $\widehat{Q}^2$-algebra. With a further change of variables
\begin{equation}
\label{eq:bsW-shift}
b^I \longrightarrow b^I + 2\sinh\eta\,p_1\,c^I \, , \quad \sigma^I \longrightarrow -\sigma^I + \tfrac12\sinh\eta\,\tilde{W}^I_1 \, , \quad \tilde{W}_1^I \longrightarrow \tilde{W}_1^I - 2\sinh\eta\,\sigma^I \, ,
\end{equation}
we bring the symbol matrix to the form:
\begin{equation}
\begin{pmatrix}
8\cosh^2\eta\,p^2 & 0 & 0 & 0 & 0 \\ 
2\sinh\eta\,p_1 & p_1 & p_2 & p_3 & p_4 \\
0 & -\cosh^2\eta\,p_2 & p_1  & -\cosh\eta\,p_4 & \cosh\eta\,p_3 \\
0 & -\cosh^2\eta\,p_3 & \cosh\eta\,p_4 & p_1 & -\cosh\eta\,p_2 \\
0 & -\cosh^2\eta\,p_4 \; & -\cosh\eta\,p_3 & \cosh\eta\,p_2 & p_1 
\end{pmatrix} \, .
\end{equation}
Observe that the shift in the anti-ghost field $b$ in \eqref{eq:bsW-shift} corresponds to the change of variables \eqref{eq:hatb} in the mode analysis of the kernel.\footnote{We could have also implemented the $(\sigma,\tilde{W}_1)$ rotation in the mode analysis of Sections \ref{sec:kerneq} and \ref{sec:cokerneq}, but elected not to do so to keep the kernel and cokernel equations in a manifestly covariant form.} Finally, taking the first line multiplied by $-p_1 \sinh\eta / (p^2 K^2)$ (which is nowhere singular) and adding it to the second line, we conclude that the relevant part of the symbol $\sigma[\,D^\mathrm{vec}_{10}\,]$ is the following $4\times4$ matrix \cite{Pestun:2007rz}:
\begin{equation}
\begin{pmatrix}
p_1 & p_2 & p_3 & p_4 \\
-\cosh^2\eta\,p_2 & p_1  & -\cosh\eta\,p_4 & \cosh\eta\,p_3 \\
-\cosh^2\eta\,p_3 & \cosh\eta\,p_4 & p_1 & -\cosh\eta\,p_2 \\
-\cosh^2\eta\,p_4 \; & -\cosh\eta\,p_3 & \cosh\eta\,p_2 & p_1 
\end{pmatrix} \, .
\end{equation}
The determinant of this matrix is
\begin{equation}
\mathrm{det}\bigl(\sigma[\,D^\mathrm{vec}_{10}\,]) = \bigl(p_1^2 + \vec{p}{\,}^2\cosh^2\eta\bigr)^2 \, ,
\end{equation}
which is nowhere vanishing provided $p_a$ is not the zero 4-vector. This shows that the symbol is invertible, and thus that the operator $D_{10}^\mathrm{vec}$ is elliptic \cite{Pestun:2007rz,Atiyah-book}.

According to \eqref{eq:gf-alg}, $\widehat{Q}^2$ acts on the spacetime manifold as a $U(1)_\varepsilon$ rotation along the Euclidean time circle, where $\varepsilon$ parametrizes the weight of the $U(1)$ action. This action has a fixed point at the origin of $\mathbb{H}_2$, located at $\eta=0$ in the coordinates \eqref{eq:NHG-metric}. At this fixed point, the symbol matrix reduces further to:
\begin{equation}
\label{eq:symbol-rel}
\begin{pmatrix}
p_1 & p_2 & p_3 & p_4 \\
-p_2 & p_1 & -p_4 & p_3 \\
-p_3 & p_4 & p_1 & -p_2 \\
-p_4 & -p_3 & p_2 & p_1 
\end{pmatrix} \, .
\end{equation}
This is also the symbol of the so-called self-dual (SD) complex \cite{Pestun:2007rz}
\begin{equation}
\label{eq:DSD}
D_{\mathrm{SD}} \; : \; \Omega^0 \stackrel{d}{\longrightarrow} \Omega^1 \stackrel{d^+}{\longrightarrow} \Omega^{2+} \, ,
\end{equation}
and therefore, at the fixed point of the $\widehat{Q}^2$ action, the equivariant index of our elliptic operator $D_{10}^\mathrm{vec}$ is captured by the equivariant index of the the elliptic complex $D_{\mathrm{SD}}$ \eqref{eq:DSD}. As a remark, we note that this is also apparent from the explicit mode analysis, where in particular the kernel equations \eqref{eq:kerneq-lambda-1} at the origin $\eta = 0$ reduce to the standard anti-self-dual (ASD) connections, 
\begin{equation}
\tilde{F}^+ = 0 \, .
\end{equation}

In typical localization calculations, one is interested in the equivariant index of such a complex when the action of $\widehat{Q}^2$ has isolated fixed points. In our situation however, the fixed locus is a codimension two submanifold: the 2-sphere sitting at the origin $\eta = 0$ of $\mathbb{H}_2$. This situation can still be efficiently dealt with by making use of the equivariant Atiyah-Singer theorem, without assuming that the set of fixed points is discrete. Namely, let $G$ be a compact Lie group acting on a smooth compact manifold $\mathcal{M}$ and let $D$ be a $G$-invariant elliptic differential operator on $\mathcal{M}$. The equivariant index of $D$ with respect to $G$ is related to the fixed point set $\mathcal{M}_g$ of $\mathcal{M}$ under $g \in G$ by (\cite{Shanahan-book}, Section 15)
\begin{equation}
\label{eq:AS-fixed}
\mathrm{ind}_g(D) = (-1)^{d_g}\,\int_{T\mathcal{M}_g}\,\mathrm{ch}_g(j^*\sigma[\,D\,])\,\frac{\mathrm{Td}(T\mathcal{M}_g^\mathbb{C})}{\mathrm{ch}_g\Bigl(\sum_r\,(-1)^r\,\bigwedge\nolimits^r\,N_g^\mathbb{C}\Bigr)}\Bigg\vert_{\mathrm{top}} \, ,
\end{equation}
where $d_g$ is the complex dimension of $\mathcal{M}_g$, $j:\mathcal{M}_g \longrightarrow \mathcal{M}$ is the inclusion mapping, $N_g$ is the normal bundle of $\mathcal{M}_g$ in $\mathcal{M}$ and $\sigma[\,D\,]$ is the symbol of $D$. The relevant characteristic classes in the above formula are the Todd class Td and the equivariant Chern character ch$_g$. Lastly, the subscript ``top'' indicates that we integrate the top-form component over the tangent space $T\mathcal{M}_g$. 
Strictly speaking, \eqref{eq:AS-fixed} is valid when the manifold $\mathcal{M}$ on which $D$ is defined is compact and without boundary. As discussed previously, we will nevertheless go ahead and use it in our setup, under the assumption that the boundary and smoothness conditions imposed on the various fields effectively make our $\mathbb{H}_2 \times S^2$ space compact.

We begin by applying the Atiyah-Singer theorem \eqref{eq:AS-fixed} to the Dolbeault operator $\bar{\partial}$ on a manifold $\mathcal{M}$. After a standard simplification of the characteristic classes on the right-hand side, we obtain \cite{Shanahan-book}
\begin{equation}
\label{eq:equiv-dolb}
\mathrm{ind}_g(\bar{\partial}) = (-1)^{d_g} \int_{\mathcal{M}_g}\,\frac{\mathrm{Td}(T\mathcal{M}_g^+)}{\mathrm{ch}_g\Bigl(\sum_r\,(-1)^r\,\bigwedge\nolimits^r\,N_g^-\Bigr)}\Bigg\vert_{\mathrm{top}} \, ,
\end{equation}
with $T\mathcal{M}_g^+$ the holomorphic tangent bundle of $\mathcal{M}_g$ and $N_g^-$ the anti-holomorphic normal bundle of $\mathcal{M}_g$ in $\mathcal{M}$. The Dolbeault complex is related to the complexified SD complex on K\"{a}hler manifolds, see e.g. \cite{Marino:1996sd}. Therefore, we can use \eqref{eq:equiv-dolb} to obtain the equivariant index of the complexified SD complex \eqref{eq:DSD} on $\mathbb{H}^2\times S^2$ with respect to the $U(1)_\varepsilon$ action, where $U(1)_\varepsilon$ acts on the neighborhood of the origin of $\mathbb{H}_2$ and leaves $S^2$ fixed:
\begin{equation}
\label{eq:equiv-SD-S2}
\mathrm{ind}_{U(1)_\varepsilon}(D^\mathbb{C}_{\mathrm{SD}})(t) = -\Bigl(\frac{1}{1 - q} + \frac{1}{1 - q^{-1}}\Bigr)\,\int_{S^2}\,\mathrm{Td}(TS^2)\big\vert_{\mathrm{top}} \, .
\end{equation}
Here $q := \exp(\mathrm{i}\,\varepsilon\,t) \in U(1)_\varepsilon$. As discussed above, \eqref{eq:equiv-SD-S2} is also the equivariant index of our differential operator $D_{10}^\mathrm{vec}$. Observe that the first term in \eqref{eq:equiv-SD-S2} corresponds to the holomorphic projection of the vector multiplet while the second term corresponds to the anti-holomorphic projection, as discussed in \cite{Pestun:2016zxk}. In the mode analysis, we have allowed modes with $n \in \mathbb{Z}^*$ for the fields of $\mathbb{X}_0^I$ and $\mathbb{X}_1^I$, and so we should keep both terms and expand each series in \eqref{eq:equiv-SD-S2} in powers of $q$ and $q^{-1}$, respectively. Doing so, we obtain
\begin{equation}
\label{eq:ind-vec-AS}
\mathrm{ind}_{U(1)_\varepsilon}(D^\mathrm{vec}_{10})(t) = - \sum_{n\geq 1}\,\mathfrak{m}\,e^{\mathrm{i}\varepsilon n t} - \sum_{n\geq 1}\,\mathfrak{m}\,e^{-\mathrm{i}\varepsilon n t} - 2  \, ,
\end{equation}
where the multiplicity is given by\footnote{We give the general definition of the multiplicity even though it evaluates to one in the case of $S^2$, in view of some generalization that will be discussed in section \ref{subsec:genus} below.}
\begin{equation}
\label{eq:multi-index}
\mathfrak{m} := \int_{S^2}\,\mathrm{Td}(TS^2)\big\vert_{\mathrm{top}} = \frac{1}{2}\,\chi(S^2) = 1 \, .
\end{equation}
The last factor of $-2$ in the index corresponds to the contribution from the zero-modes with $n=0$. As discussed in \ref{sec:mode-result}, these are the constant modes of the ghost and anti-ghost fields, and we discard them from the spectrum. Finally, according to \eqref{eq:H-eigen-2} we set $\varepsilon = 2\,v_1{}^{-1/2}$ and use the rule \eqref{eq:1-loop-formal} to read off the determinant:
\begin{equation}
\label{eq:Z-index}
Z^{\mathrm{vec}}_{\textnormal{1-loop}} = \prod_{n \geq 1} \Bigl(\frac{2\mathrm{i}n}{\sqrt{v_1}}\Bigr)^{\mathfrak{m}/2}\prod_{n \geq 1} \Bigl(-\frac{2\mathrm{i}n}{\sqrt{v_1}}\Bigr)^{\mathfrak{m}/2} = \prod_{n \geq 1} \Bigl(\frac{4n^2}{v_1}\Bigr)^{1/2} \, .
\end{equation}
This result is in agreement with the explicit mode computation \eqref{eq:Z-modes}.

\subsection{Method III: Refinement and Atiyah-Bott fixed point theorem}
\label{sec:FP-result}

A third way to obtain the one-loop determinant is to introduce a refinement of the $\widehat{Q}^2$ Hamiltonian, exactly as was done in the field theory analogue for the topologically twisted index \cite{Benini:2015noa}. Loosely speaking, this refinement mimics the $\Omega$-background of Nekrasov \cite{Nekrasov:2002qd}. The possibility of turning on such a refinement can be justified by the existence of \emph{rotating} supersymmetric black holes in gauged supergravity \cite{Hristov:2018spe} that generalize the static near-horizon geometries considered in the present paper. These rotating solutions admit a smooth limit back to the unrefined, i.e.\ static, case. Here we can use this in order to take an alternative route in calculating the one-loop determinant. At a hands-on level, we can build a refined Hamiltonian such that its set of fixed points is isolated, as opposed to the $S^2$ case discussed in the previous section. We can then apply the standard Atiyah-Bott fixed point theorem \cite{Atiyah-book} to compute the index of the operator $D_{10}$ associated to the refined Hamiltonian, and take a suitable unrefined limit at the end.

Consider then a deformation of the $\widehat{Q}^2$ operator \eqref{eq:gf-alg} defined by
\begin{equation}
H_{\varepsilon_1,\varepsilon_2} := \mathcal{L}_{v(\epsilon_1,\epsilon_2)} + \delta_{\rm R}(\Lambda^i{}_j) \, , \quad \text{where} \quad v^\mu(\epsilon_1,\epsilon_2) = \bigl(\epsilon_1\,,0\,,0\,,\epsilon_2\bigr)^\mathrm{T} \, .
\end{equation}
Compared to \eqref{eq:H-eigen}, the eigenvalues of $H_{\varepsilon_1,\varepsilon_2}$ are labeled by $\mathfrak{n} = \{n_1,n_2\} \in \mathbb{Z}^2$,
\begin{equation}
\label{eq:refined-H}
H_{\varepsilon_1,\varepsilon_2}\mathbb{X}^I_{0,1} = \bigl(\mathrm{i}\varepsilon_1\,n_1 + \mathrm{i}\varepsilon_2\,n_2\bigr)\,\mathbb{X}^I_{0,1}  \, .
\end{equation}
Using hyperbolic-stereographic coordinates for $\mathbb{H}_2 \times S^2$, 
\begin{equation}
w = \tanh\tfrac{\eta}{2}\,e^{\mathrm{i}\tau} \, , \quad z = \tan\tfrac{\theta}{2}\,e^{\mathrm{i}\varphi} \, ,
\end{equation}
it is clear that $H_{\varepsilon_1,\varepsilon_2}$ generates a $U(1)_{\varepsilon_1} \times U(1)_{\varepsilon_2}$ action,
\begin{equation}
\label{eq:refined-spacetime-action}
(w,z) \;\; \stackrel{e^{tH_{\varepsilon_1,\varepsilon_2}}}{\longmapsto} \;\; \bigl(q_1 w,q_2 z\bigr) \, , \quad \text{where} \quad q_1 := \exp(\mathrm{i}\varepsilon_1 t) \, , \;\; q_2 := \exp(\mathrm{i}\varepsilon_2 t) \, .
\end{equation}
The $\mathbb{H}_2 \times S^2$ space has two isolated fixed points under the action of $e^{tH_{\varepsilon_1,\varepsilon_2}}$, given by $(w=0,z=0)$ and $(w=0,z^{-1}=0)$ and corresponding to the North Pole (NP) and South Pole (SP) of the 2-sphere sitting at the origin of $\mathbb{H}_2$, respectively. We can therefore use the Atiyah-Bott fixed point theorem\footnote{Modulo the assumption that the differential operator in the refined case is transversally elliptic with respect to the $U(1)_{\varepsilon_1} \times U(1)_{\varepsilon_2}$ action, and the non-compactness issues that we have already discussed.}, to compute the equivariant index of the refined $D^\mathrm{vec}_{10}$ operator under the $U(1)_{\varepsilon_1} \times U(1)_{\varepsilon_2}$ action,
\begin{equation}
\mathrm{ind}_{U(1)_{\varepsilon_1} \times U(1)_{\varepsilon_2}}(D_{10}^\mathrm{vec})(t) = \sum_{x\,|\,\widetilde{x}=x}\,\frac{\mathrm{Tr}_{\mathbb{X}^I_0,\mathbb{X}^I_1} (-1)^F\,e^{tH_{\varepsilon_1,\varepsilon_2}}}{\mathrm{det}(1-\partial\widetilde{x}/\partial x)} \, , \quad \text{where} \quad \widetilde{x} = e^{tH_{\varepsilon_1,\varepsilon_2}}\,x \, .
\end{equation}
According to \eqref{eq:refined-spacetime-action}, at each fixed point the factor at the denominator is the product
\begin{equation}
\label{eq:AB-denom}
\mathrm{det}(1-\partial\widetilde{x}/\partial x) = (1-q_1)(1-q_1^{-1})(1-q_2)(1-q_2^{-1}) \, .
\end{equation}
To compute the traces in the numerator, we note that locally the fixed points look like ${\cal R}^4$ with an associated $SO(4) \sim SU(2)_+ \times SU(2)_-$ symmetry. The planes labeled by $w$ and $z$ rotate under this $SO(4)$ depending on the fixed point. For the bosonic fields $\mathbb{X}^I_0$, the scalar $\sigma$ is neutral under the $SO(4)$, while the vector field has two components $(\tilde{W}_w,\tilde{W}_{\bar{w}})$ rotating with charge $(-1,+1)$ and weight $q_1$, and two components $(\tilde{W}_z,\tilde{W}_{\bar{z}})$ rotating with charge $(-1,+1)$ and weight $q_2$. So, both at the NP and SP fixed points,
\begin{equation}
\mathrm{Tr}_{\mathbb{X}^I_0} (-1)^F\,e^{tH_{\varepsilon_1,\varepsilon_2}} = 1 + q_1^{-1} + q_1 + q_2^{-1} + q_2 \, .
\end{equation}
For the fermionic fields $\mathbb{X}^I_1$, the ghosts are scalars and neutral under the $SO(4)$, while the fermions $\lambda^{ij}$ rotate depending on their SU(2)$_{\rm R}$ components. A basis to expand the fermions at the NP is given by \cite{Gupta:2015gga}
\begin{equation}
\lambda^{11} \sim \frac{\partial}{\partial w} \wedge \frac{\partial}{\partial z} \, , \quad \lambda^{12} \sim \frac{\partial}{\partial w} \wedge \frac{\partial}{\partial \bar{w}} + \frac{\partial}{\partial z} \wedge \frac{\partial}{\partial \bar{z}} \, , \quad \lambda^{22} \sim \frac{\partial}{\partial \bar{w}} \wedge \frac{\partial}{\partial \bar{z}} \, .
\end{equation}
Using \eqref{eq:refined-spacetime-action} we then obtain the charges of $\lambda^{ij}$, which leads to 
the NP fermion trace
\begin{equation}
\label{eq:refined-fermion-trace}
\mathrm{Tr}_{\mathbb{X}^I_1} (-1)^F\,e^{tH_{\varepsilon_1,\varepsilon_2}} = - 2 - q_1^{-1} q_2^{-1} - 1 - q_1 q_2 \, .
\end{equation}
At the SP, the basis to expand $\lambda^{ij}$ is given in terms of the $w$ and $u := z^{-1}$ coordinates,
\begin{equation}
\lambda^{11} \sim \frac{\partial}{\partial w} \wedge \frac{\partial}{\partial \bar{u}} \, , \quad \lambda^{12} \sim \frac{\partial}{\partial w} \wedge \frac{\partial}{\partial \bar{w}} - \frac{\partial}{\partial u} \wedge \frac{\partial}{\partial \bar{u}} \, , \quad \lambda^{22} \sim \frac{\partial}{\partial \bar{w}} \wedge \frac{\partial}{\partial u} \, .
\end{equation}
This shows that the SP fermion trace is also given by \eqref{eq:refined-fermion-trace}. Together with the denominator \eqref{eq:AB-denom}, each fixed point gives a contribution to the index of
\begin{equation}
\label{eq:refined-FP-index}
-\frac{1 + q_1q_2}{(1-q_1)(1-q_2)} \, .
\end{equation}
This contribution should be expanded in either positive or negative powers of $(q_1,q_2)$ at the NP and SP. We write the total index as
\begin{equation}
\label{eq:refined-index}
\mathrm{ind}_{U(1)_{\varepsilon_1} \times U(1)_{\varepsilon_2}}(D_{10}^\mathrm{vec})(t) = -\Bigl[\frac{1+q_1 q_2}{(1-q_1)(1-q_2)}\Bigr]_{\mathrm{NP}} - \Bigl[\frac{1+ q_1 q_2}{(1-q_1)(1-q_2)}\Bigr]_{\mathrm{SP}} \, .
\end{equation}

We are interested in the limit where $\varepsilon_2 \rightarrow 0$ ($q_2 \rightarrow 1$) of the above index. In this limit, the action of the refined Hamiltonian \eqref{eq:refined-H} reduces to the $U(1)_{\varepsilon_1}$ of Sections \ref{sec:index-result} and the quantum number $n_1 = n$ associated with $q_1$ corresponds to the mode decomposition along the Euclidean time circle. In Section \ref{sec:index-result} we explained how we should keep all modes with $n \in \mathbb{Z}^*$, and this corresponds to expanding the refined index \eqref{eq:refined-index} in positive powers of $q_1$ at the NP and negative powers of $q_1$ at the SP. We now choose a $q_2$-expansion before taking the unrefined limit. For the NP fixed point we expand in positive powers:
\begin{equation}
\label{eq:NP-good-exp}
-\Bigl[\frac{1+q_1 q_2}{(1-q_1)(1-q_2)}\Bigr]_{\mathrm{NP}} = - \sum_{n_1,n_2\geq 0}\,(1+q_1 q_2)\,q_1^{n_1}\,q_2^{n_2} \, ,
\end{equation}
while for the SP fixed point we expand in negative powers,
\begin{equation}
\label{eq:SP-good-exp}
-\Bigl[\frac{1+q_1 q_2}{(1-q_1)(1-q_2)}\Bigr]_{\mathrm{SP}} = - \sum_{n_1,n_2\geq0}\,\bigl(1+ q_1^{-1}q_2^{-1}\bigr)\,q_1^{-n_1}\,q_2^{-n_2} \, .
\end{equation}
Then, taking the limit $\varepsilon_2 \rightarrow 0$, we obtain the contribution from the NP fixed point
\begin{equation}
\lim_{\varepsilon_2\rightarrow 0}\;\sum_{n_1,n_2\geq0}\,-\bigl(1 + q_1 q_2\bigr)\,q_1^{n_1}\,q_2^{n_2} = -\Bigl(\sum_{n_2\geq 0}\,1\Bigr)\,\sum_{n\geq 0} (1+q_1)\,q_1^n \, .
\end{equation}
The divergent prefactor is due to the first order pole of the index in the unrefined limit. We use zeta-function regularization\footnote{We interpret the sum as $\sum_{n\geq 1} n^0 = \zeta_\mathrm{R}(0) = -1/2$ where the last equality uses analytic continuation.} to write the right-hand side above as
\begin{equation}
\lim_{\varepsilon_2\rightarrow 0}\;\sum_{n_1,n_2\geq0}\,-\bigl(1 + q_1 q_2\bigr)\,q_1^{n_1}\,q_2^{n_2} = -\bigl(1+\zeta_{\mathrm{R}}(0)\bigr)\,\Bigl(2\,\sum_{n\geq 1}\,q_1^n + 1\Bigr) \, ,
\end{equation}
where we split off the $n=0$ term in the series. As discussed previously, this term corresponds to the constant ghost and anti-ghost zero-mode, which we discard from the spectrum. Thus, in the unrefined limit, the choice of expansion \eqref{eq:NP-good-exp} yields a contribution from the NP fixed point to the index of
\begin{equation}
\mathrm{NP} \;\; : \quad -\sum_{n \geq 1}\,q_1^n \, .
\end{equation}
Similarly, using \eqref{eq:SP-good-exp} at the SP and zeta-function regularization for the limit $\varepsilon_2 \rightarrow 0$, we obtain a contribution to the index from the SP of
\begin{equation}
\mathrm{SP} \;\; : \quad  -\sum_{n \geq 1}\,q_1^{-n} \, .
\end{equation}
Using the rule \eqref{eq:1-loop-formal}, we then obtain the one-loop determinant in the unrefined limit from the Atiyah-Bott fixed point theorem upon setting $\varepsilon_1 = 2\,v_1{}^{-1/2}$,
\begin{equation}
Z_{\text{1-loop}}^\text{vec} = \prod_{n \geq 1} \Bigl(\frac{2\mathrm{i}n}{\sqrt{v_1}}\Bigr)^{1/2}\prod_{n \geq 1} \Bigl(-\frac{2\mathrm{i}n}{\sqrt{v_1}}\Bigr)^{1/2} = \prod_{n \geq 1} \Bigl(\frac{4n^2}{v_1}\Bigr)^{1/2} \, .
\end{equation}
This shows that our choice of $q_2$-expansion at the NP and SP reproduces the results obtained using the mode analysis in section \ref{sec:mode-result} and the more general form of the Atiyah-Singer theorem when the fixed points are not isolated in section \ref{sec:index-result}.

\subsection{Regularization and scale-invariant form}
\label{sec:scale-inv}

Regardless of the method used to compute it, we come to the conclusion that the one-loop determinant for a given vector multiplet takes the form of an infinite product,
\begin{equation}
Z^\mathrm{vec}_{\textnormal{1-loop}} = \prod_{n\geq 1}\,\Bigl(\frac{4n^2}{v_1}\Bigr)^{1/2} \, .
\end{equation}
To regularize this expression, we use zeta-function regularization:
\begin{equation}
\log Z^{\mathrm{vec}}_{\textnormal{1-loop}} = \frac12\sum_{n\geq1}\bigl[\,\log(4 n^2) - \log v_1\,\bigr] = - \frac12\log(2) + \frac12\log(2\pi) +  \frac14\log v_1 \, .
\end{equation}
Dropping the purely numerical constants, we finally obtain the one-loop determinant for a given vector multiplet:
\begin{equation}
\label{eq:Z-reg}
Z^{\mathrm{vec}}_{\textnormal{1-loop}} = v_1{}^{1/4} \, .
\end{equation}
At this stage, the quantity $v_1$ controlling the one-loop determinant is a constant parameter for the size of the $\mathbb{H}_2$ space, although it is subject to scaling transformations. To work in terms of scale-invariant quantities in the superconformal framework, we should use the Einstein frame metric
\begin{equation}
G_{\mu\nu} = \mathring{g}_{\mu\nu}\,\chi_\mathrm{V}(X_+,X_-) \, ,
\end{equation}
which depends explicitly on the scalar fields of the vector multiplets through the K\"{a}hler potential $\chi_\mathrm{V}$ (see appendix \ref{app:A} for definitions). Since $w(\mathring{g}_{\mu\nu}) = -2$ and $w(\chi_\mathrm{V}) = 2$, the $G$-metric has indeed zero Weyl weight. In terms of the $\mathring{g}$-metric, the $\widehat{Q}^2$ Hamiltonian controlling the one-loop determinant had Weyl weight $w(H) = 1$. We can build a scale-invariant Hamiltonian $\widetilde{H}$ by multiplying $H$ by appropriate factors of $\chi_\mathrm{V}$, namely
\begin{equation}
\widetilde{H} := \chi_\mathrm{V}^{-1/2}\,H \, .
\end{equation}
The $\widetilde{H}$ eigenvalues are
\begin{equation}
\widetilde{H}\,\mathbb{X}^I_{0,1} = 2\mathrm{i}n\,\bigl(v_1\,\chi_\mathrm{V}\bigr)^{-1/2}\,\mathbb{X}^I_{0,1} \, .
\end{equation}
The multiplicities of these eigenvalues are pure numbers and are not modified compared to our previous computations with the Hamiltonian $H$. So using \eqref{eq:1-loop-formal} 
and zeta-function regularization gives a one-loop determinant which depends explicitly on the scalar fields of the vector multiplet,
\begin{equation}
\label{eq:vec-one-loop-final}
Z^\mathrm{vec}_{\text{1-loop}}(X_+,X_-) = \bigl(v_1\,\chi_\mathrm{V}(X_+,X_-)\bigr)^{1/4} \, .
\end{equation}
Note that this is now written in a scale-invariant form, since $w(v_1\,\chi_\mathrm{V}) = 0$.

\subsection{Generalization to higher genus}
\label{subsec:genus}

The index theorem of section \ref{sec:index-result} is particularly suited to discuss generalizations of our result when the horizon has a more general topology, such as a higher genus Riemann surface $\Sigma_{\rm g}$ (see \cite{Benini:2016hjo,Closset:2016arn} for the analogous calculation in field theory). In this case, the fixed codimension two submanifold under the $\widehat{Q}^2$ action in \eqref{eq:AS-fixed} will be the surface $\Sigma_{\rm g}$, and repeating the steps leading to \eqref{eq:Z-index} will yield the same form of the one-loop determinant where the multiplicity of eigenvalues $\mathfrak{m}$ defined in \eqref{eq:multi-index} is replaced by the integral of the Todd class of $T\Sigma_{\rm g}$ over the Riemann surface. Thus, we expect that in this situation,
\begin{equation}
Z^{\mathrm{vec},\,\Sigma_{\rm g}}_{\textnormal{1-loop}} = \prod_{n \geq 1} \Bigl(\frac{4n^2}{v_1}\Bigr)^{\chi(\Sigma_{\rm g})/4} \, ,
\end{equation}
where $\chi(\Sigma_{\rm g})$ is the Euler characteristic of the Riemann surface. After zeta-function regularization and in the scale-invariant formulation discussed in section \ref{sec:scale-inv}, we will then obtain
\begin{equation}
Z^{\mathrm{vec},\,\Sigma_{\rm g}}_{\textnormal{1-loop}}(X_+,X_-) = \bigl(v_1\,\chi_\mathrm{V}(X_+,X_-)\bigr)^{\chi(\Sigma_{\rm g})/8} \, ,
\end{equation}
for each vector multiplet. It is worth emphasizing that this generalization to other horizon topology is easily derived from the index theorem, while it would require repeating the mode analysis which relied on expanding fields along the appropriate harmonics. This illustrates the power of the Atiyah-Singer theorem when computing the determinants arising in a typical localization computation. Note that the explicit mode analysis is still helpful to discuss the precise choice of boundary conditions on the various fields and identify potential zero-modes, which are important aspects one needs to deal with on non-compact spaces.

\section{Hypermultiplets}
\label{sec:hypers}

So far we have discussed the one-loop determinant for a generic abelian vector multiplet. We now want to consider the one-loop determinant for a hypermultiplet. There are two types of hypermultiplets that will be relevant. The first is the \emph{compensating} hypermultiplet, which is needed to ensure that the superconformal theory used in localization is gauge-equivalent to the usual Poincar\'{e} gauged supergravity \cite{deWit:1980lyi}. The second is a generic physical hypermultiplet. As we will explain, the two require separate treatment as we argue that they satisfy different reality conditions leading to different one-loop determinant contributions.

\subsection{The compensating hypermultiplet}

As for the vector multiplet discussed in section \ref{sec:vectors}, we should identify the relevant differential operator $D_{10}$ to compute the one-loop determinant of the compensating hypermultiplet. For this it is important to note that, due to the gauging in superconformal gravity, the compensating hypermultiplet couples to a special linear combination of vector multiplets specified by the FI parameters $\xi_I$.
We will denote this linear combination by $\xi_I \mathbb{V}^I$. To be explicit, recall that the off-shell transformation rules of the compensating hypermultiplet fields under the localizing supercharge $\widehat{Q}$ are \cite{Hristov:2018lod,deWit:2017cle}
\begin{align}
\widehat{Q}A_i{}^\alpha = &\; 2\,\bar{\xi}_i\,\gamma^5\,\zeta^\alpha + \xi_I c^I\,t^\alpha{}_\beta A_i{}^\beta \, , \nonumber\\[1mm]
\widehat{Q}\zeta^\alpha =&\; -\mathrm{i}\,\Slash{\mathcal{D}} A_i{}^\alpha\,\xi^i - 2\mathrm{i}\,\xi_I \sigma^I\,t^\alpha{}_\beta A_i{}^\beta \xi^i + 2\,\xi_I \rho^I\,t^\alpha{}_\beta A_i{}^\beta \gamma^5\xi^i + H_i{}^\alpha\check{\xi}^i + A_i{}^\alpha \kappa^i - \xi_I c^I\,t^\alpha{}_\beta \zeta^\beta \, , \nonumber \\[1mm]
\widehat{Q}H_i{}^\alpha =&\; \mathrm{i}\,\bar{\check{\xi}}_i \gamma^5 \Slash{\mathcal{D}}\zeta^\alpha +  \xi_I c^I\,t^\alpha{}_\beta H_i{}^\beta \, ,
\end{align}
where the gauging generators are anti-Hermitian, $\bigl(t^\alpha{}_\beta\bigr)^\dagger := t_\alpha{}^\beta = -t^\beta{}_\alpha = \Omega^{\beta\gamma}\,t_\gamma{}^\delta\,\Omega_{\delta\alpha}$, the derivative $\mathcal{D}_\mu$ is covariantized with respect to the abelian gauge symmetry of the linear combination of vector multiplets, e.g.
\begin{equation}
\mathcal{D}_\mu A_i{}^\alpha = \partial_\mu A_i{}^\alpha + \tfrac12\,\mathcal{V}_{\mu\,i}{}^j A_j{}^\alpha - \mathrm{i}\,\xi_I \tilde{W}_\mu^I\,t^\alpha{}_\beta\,A_i{}^\beta \, ,
\end{equation}
and the constrained parameters $\check{\xi}^i$ are auxiliary symplectic-Majorana spinors required to satisfy
\begin{equation}
\bar{\xi}_{i+}\check{\xi}^j_+ = \bar{\xi}_{i-}\check{\xi}^j_- \, , \qquad \bar{\check{\xi}}_{i\pm}\check{\xi}^j_\pm = \bar{\xi}_{i\mp}\xi^j_\mp \, , \qquad \bar{\check{\xi}}_{i\pm}\gamma^\mu\check{\xi}^j_\mp = \bar{\xi}_{i\pm}\gamma^\mu\xi^j_\mp \, .
\end{equation}
Above, the subscripts denote chiral projections. Note that in this formulation we have introduced the scalar auxiliary fields $H_i{}^\alpha$ together with the constrained parameters $\check{\xi}^i$ in order to close the algebra of the supercharge $\widehat{Q}$ off-shell according to \eqref{eq:gf-alg}. This is suited for localization, and is explained in more detail in \cite{Hristov:2018lod}. 

Just as in the vector multiplet computation, we can introduce the twisted hyperini
\begin{equation}
\lambda_i{}^\alpha := 2\,\bar{\xi}_i\,\gamma^5\,\zeta^\alpha \, , \quad \mathrm{and} \quad \Xi_i{}^\alpha := \bar{\check{\xi}}_i\,\zeta^\alpha \, ,
\end{equation}
in terms of which the original hyperini are given by
\begin{equation}
\zeta^\alpha = K^{-1}\bigl(\gamma^5\,\xi^i\,\lambda_i{}^\alpha + 2\,\check{\xi}^i\,\Xi_i{}^\alpha\bigr) \, .
\end{equation}
The transformation rules under $\widehat{Q}$ now take the following form:
\begin{align}
\label{eq:Qhat-transfo-comp}
\widehat{Q}A_i{}^\alpha =&\; \lambda_i{}^\alpha + \xi_I c^I\,t^\alpha{}_\beta\,A_i{}^\beta \, , \nonumber \\[1mm]
\widehat{Q}\lambda_i{}^\alpha =&\; \mathring{v}^\mu \mathcal{D}_\mu A_i{}^\alpha + \tfrac{\mathrm{i}}{\sqrt{v_1}}\,\sigma_{3\,i}{}^j\,A_j{}^\alpha + 2\,\xi_I(K\,\rho^I + 2\mathrm{i}\,\sigma^I)\,t^\alpha{}_\beta\,A_i{}^\beta - \xi_Ic^I\,t^\alpha{}_\beta\,\lambda_i{}^\beta \, , \\[1mm]
\!\!\widehat{Q}\Xi_i{}^\alpha =&\; \tfrac12\,KH_i{}^\alpha - \check{K}^\mu{}_i{}^j\mathcal{D}_\mu A_j{}^\alpha + \tfrac{\mathrm{i}}{2\sqrt{v_1}}\sinh\eta\,\sigma_{3\,i}{}^j A_j{}^\alpha - 2\mathrm{i}\sinh\eta\,\xi_I\sigma^I\,t^\alpha{}_\beta A_i{}^\beta - \xi_Ic^I\,t^\alpha{}_\beta\,\Xi_i{}^\beta \, , \nonumber
\end{align}
together with the transformation of $H_i{}^\alpha$ that we will not need explicitly. Above, we have defined the pseudo-real bilinear
\begin{equation}
\check{K}^\mu{}_i{}^j := \mathrm{i}\,\bar{\check{\xi}}_i\,\gamma^\mu\,\xi^j \, .
\end{equation}

The transformations \eqref{eq:Qhat-transfo-comp} make it clear that the compensating hypermultiplet couples to the linear combination of vector multiplets $\xi_I\mathbb{V}^I$, which includes the corresponding ghost fields needed to fix the abelian gauge symmetry in the path-integral. Because of this coupling, in order to correctly identify the $D^\mathrm{comp}_{10}$ operator relevant for the one-loop determinant, we must build a fermionic deformation $\widehat{\mathcal{V}}^{\,\mathrm{comp}}$ out of an extended multiplet comprising the twisted hyperini and the relevant linear combination of twisted gaugini. The equivariant index of $D^\mathrm{comp}_{10}$ will encode the contributions to the one-loop determinant coming from the compensating hypermultiplet and from the vector multiplet $\xi_I\mathbb{V}^I$. We use the following fermionic deformation
\begin{align}
\begin{split}
\label{eq:comp-deform}
\!\!\!\!\widehat{\mathcal{V}}^{\,\textnormal{comp}} = \int d^4x\,\frac{\sqrt{\mathring{g}}}{K^2}\,\Bigl[&\xi_I\xi_J\Bigl\{\lambda^I(\widehat{Q}\lambda^J)^\dagger + \lambda^{I\,\mu}(\widehat{Q}\lambda^J_\mu)^\dagger + 2\,\lambda^{I\,ij}(\widehat{Q}\lambda^{J\,ij})^\dagger +  K^2\,b^I G(\tilde{W}^J)\Bigr\} \\
&+ \lambda_i{}^\alpha\bigl(\widehat{Q}\lambda_i{}^\alpha\bigr)^\dagger + 4\,\Xi_i{}^\alpha \bigl(\widehat{Q}\Xi_i{}^\alpha\bigr)^\dagger\Bigr]\Big\vert_{\textnormal{quad.}} \, ,
\end{split}
\end{align}
and only retain terms of quadratic order in the fields to compute the one-loop determinant. We also use the following cohomological split \cite{Murthy:2015yfa,Jeon:2018kec},
\begin{equation}
\mathbb{X}^\mathrm{comp}_0 = \{\xi_I\sigma^I,\,\xi_I\tilde{W}_\mu^I,\,A_i{}^\alpha\} \, , \qquad \mathbb{X}^\mathrm{comp}_1 = \{\xi_I\lambda^{ij\,I},\,\xi_Ic^I,\,\xi_Ib^I,\,\Xi_i{}^\alpha\} \, .
\end{equation}
The first line in $\widehat{\mathcal{V}}^{\,\textnormal{comp}}$ contributes a term similar to \eqref{eq:D10} contracted with two FI parameters. To obtain the contribution from the second line, we write explicitly
\begin{align}
\begin{split}
\label{eq:lambda-comp-coho}
(\widehat{Q}\lambda_i{}^\alpha)^\dagger =&\; \mathring{v}^\mu\partial_\mu(A_i{}^\alpha)^\dagger + 2\mathrm{i}\,\xi_I\,\bigl(\mathring{v}^\mu\tilde{W}_\mu^I - 4\,\sigma^I + \tfrac{\mathrm{i}}{2}\,\widehat{Q}c^I\bigr)\,(t^\alpha{}_\beta A_i{}^\beta)^\dagger \\
&\;+ \tfrac{\mathrm{i}}{\sqrt{v_1}}\,\varepsilon^{ik}\varepsilon_{jl}\,\sigma_{3\,k}{}^l\,(A_j{}^\alpha)^\dagger - \xi_I c^I\,(t^\alpha{}_\beta\,\widehat{Q}A_i{}^\beta)^\dagger \, ,
\end{split}
\end{align}
where we used the reality conditions \eqref{eq:rot-real-cond}, the fact that $\mathring{v}^\mu\mathcal{V}_\mu{}^i{}_j = 0$ on the half-BPS background (see \eqref{eq:1/2-BPS-SU(2)-connection}), and the fact that the ghost field $c^I$ is anti-commuting. We must also specify the reality conditions on the scalars $A_i{}^\alpha$. In keeping with the choice of contour for the vector multiplet fields \eqref{eq:rot-real-cond}, we should use pseudo-imaginary sections,
\begin{equation}
\label{eq:sec-real-cond-phys}
(A_i{}^{\alpha})^\dagger = -\varepsilon^{ij}\Omega_{\alpha\beta}\,A_j{}^{\beta} \, .
\end{equation}
However, we must recall that we are dealing with a compensating multiplet in this section. In general, such compensating multiplets appear in the action with a ``wrong sign'' kinetic term and do not carry any physical degrees of freedom. In order to take this into account, we will instead use a real contour,
\begin{equation}
\label{eq:sec-real-cond-comp}
(A_i{}^{\alpha\,{\rm comp}})^\dagger = \varepsilon^{ij}\Omega_{\alpha\beta}\,A_j{}^{\beta\,{\rm comp}} \, ,
\end{equation}
when computing the contribution of the compensating hypermultiplet to the one-loop determinant.\footnote{Note that in contrast, the usual vector multiplet compensator of the conformal supergravity formalism must be treated on the same footing as the other $n_V$ vector multiplets due to our choice of gauge-fixing for the dilatation symmetry, as already discussed in \cite{Hristov:2018lod}.}

We can now find the contribution to $D^\mathrm{comp}_{10}$ from the product $\lambda_i{}^\alpha\bigl(\widehat{Q}\lambda_i{}^\alpha\bigr)^\dagger$. Because of the gauging, this product contains terms up to fourth order in the fields of $\mathbb{X}_0^\mathrm{comp}$ and $\mathbb{X}_1^\mathrm{comp}$. To focus on the quadratic terms we must set some of the fields to their localizing expectation values. We will denote these values by a subscript ``loc'' to distinguish them from the fluctuations around the localization locus. Moreover, being primarily interested in the symbol of the differential operator $D_{10}^\mathrm{comp}$, we will only retain quadratic terms that contain at least one derivative. This way, the only contribution to $\sigma[\,D_{10}^\mathrm{comp}\,]$ from the $\lambda_i{}^\alpha\bigl(\widehat{Q}\lambda_i{}^\alpha\bigr)^\dagger$ term in $\widehat{\mathcal{V}}^{\,\mathrm{comp}}$ is
\begin{equation}
\lambda_i{}^\alpha\bigl(\widehat{Q}\lambda_i{}^\alpha\bigr)^\dagger\; \ni \; \mathrm{i}\,\chi_\mathrm{H}^{1/2}\,\sigma_3{}^i{}_j\,\xi_I c^I\,\mathring{v}^\mu\partial_\mu A_i{}^j \, .
\end{equation}
Above, we have used that the localizing expectation value of the scalars $A_i{}^\alpha$ are equal to their on-shell values \eqref{eq:1/2-BPS-hyp} as shown in \cite{Hristov:2018lod},
\begin{equation}
A_i{}^\alpha|_{\textnormal{loc}} = \chi_{\mathrm{H}}^{1/2}\,\delta_i{}^\alpha \, .
\end{equation}

The contribution to the operator $D^\mathrm{comp}_{10}$ from the $\Xi_i{}^\alpha \bigl(\widehat{Q}\Xi_i{}^\alpha\bigr)^\dagger$ term is obtained using the reality conditions \eqref{eq:rot-real-cond}, \eqref{eq:sec-real-cond-comp} and reads
\begin{equation}
\Xi_i{}^\alpha\bigl(\widehat{Q}\Xi_i{}^\alpha\bigr)^\dagger = \Xi_i{}^\alpha\,\varepsilon^{ij}\Omega_{\alpha\beta}\Bigl[\widehat{Q}\Xi_j{}^\beta - 2\mathrm{i}\,\check{K}^\mu{}_j{}^k\,\xi_I \tilde{W}^I_\mu\,t^\beta{}_\gamma\,A_k{}^\gamma + 4\mathrm{i}\sinh\eta\,\xi_I \sigma^I\,t^\beta{}_\gamma\,A_j{}^\gamma\Bigr] \, .
\end{equation}
This term does not contain any derivatives, and so the term $\Xi_i{}^\alpha \bigl(\widehat{Q}\Xi_i{}^\alpha\bigr)^\dagger$ in \eqref{eq:comp-deform} does not contribute to the symbol of $D_{10}^\mathrm{comp}$.\footnote{This statement relies on the choice of reality condition for the compensating hypermultiplet scalars given \eqref{eq:sec-real-cond-comp}. Below we will discuss a different choice for physical hypermultiplets.}

Putting the contributions together, we obtain the symbol $\sigma[\,D_{10}^\mathrm{comp}\,]$. We recover the symbol $\sigma[\,D_{10}^\mathrm{vec}\,]$ in the vector multiplet sector $\xi_I\mathbb{V}^I$, augmented by a single term:
\begin{align}
\begin{split}
\mathbb{X}_1^{\mathrm{comp}}\sigma[&\,D_{10}^\mathrm{comp}\,]\mathbb{X}_0^{\mathrm{comp}} = \\
&\xi_I\mathbb{X}_1^{I\,\mathrm{vec}}\sigma[\,D_{10}^\mathrm{vec}\,]\,\xi_I\mathbb{X}_0^{I\,\mathrm{vec}} + \xi_Ic^I\bigl(2\,\chi_\mathrm{H}^{1/2}\sinh\eta\,p_1\bigr)\,\mathrm{i}\,(A_1{}^1 - A_2{}^2) \, .
\end{split}
\end{align}
Observe that at the origin of $\mathbb{H}_2$, the additional term coming from the compensating hypermultiplet fields vanishes. Therefore, the coupling between the compensating hypermultiplet and the vector multiplet $\xi_I\mathbb{V}^I$ effectively vanishes at the fixed point of the $\widehat{Q}^2$ action, and the one-loop determinant contribution from the extended multiplet sector considered here simply reduces to a factor of \eqref{eq:vec-one-loop-final} for the linear combination $\xi_I\mathbb{V}^I$. In conclusion, we can use the result \eqref{eq:vec-one-loop-final} for the contribution of both a generic vector multiplet and the combination $\xi_I\mathbb{V}^I$, while the compensating hypermultiplet gives a trivial contribution.

\subsection{Physical hypermultiplets}

The superconformal formalism also allows us to consider physical hypermultiplets in the black hole near-horizon background. For one such generic multiplet, the difference is that the fields in the physical hypermultiplet potentially couple only to physical vector multiplets such that the multiplet survives the gauge-fixing to Poincar\'{e} supergravity. Below we will consider the case where there is no explicit coupling between the physical hypermultiplets and the rest of the Lagrangian, but in the end we can actually show that the final result remains unchanged with an arbitrary coupling. It is therefore enough to consider a fermionic deformation (here ${\rm P} = 1 \ldots n_H$ labels the physical hypermultiplet of interest)
\begin{equation}
\label{eq:hyp-deform}
\widehat{\mathcal{V}}^{\textnormal{hyp}} = \int d^4x\,\frac{\sqrt{\mathring{g}}}{K^2}\,\sum_{{\rm P}}\Bigl[\lambda_i{}^{\alpha\,{\rm P}}\bigl(\widehat{Q}\lambda_i{}^{\alpha\,{\rm P}}\bigr)^\dagger + 4\,\Xi_i{}^{\alpha\,{\rm P}} \bigl(\widehat{Q}\Xi_i{}^{\alpha\,{\rm P}}\bigr)^\dagger\Bigr] \, ,
\end{equation}
to obtain the contribution to the one-loop determinant from $n_H$ physical hypermultiplet. In this section we will once again omit the label ${\rm P}$ on the fields for convenience. It is however important to remember that we are considering a physical hypermultiplet and not the conformal compensator, for the following reason. As we showed in the previous subsection, when the scalars $A_i{}^\alpha$ are pseudo-real there are no derivative terms contributing to $\widehat{\mathcal{V}}^{\textnormal{hyp}}$. When considering a physical hypermultiplet however, we will use the contour \eqref{eq:sec-real-cond-phys} and take the scalar sections to be pseudo-imaginary.
We then have the following action of the Hermitian conjugate on $\widehat{Q}\Xi_i{}^\alpha$:
\begin{equation}
\bigl(\widehat{Q}\Xi_i{}^{\alpha}\bigr)^\dagger = \varepsilon^{ij}\Omega_{\alpha\beta}\Bigl[\widehat{Q}\Xi_j{}^{\beta} + 2\,\check{K}^\mu{}_j{}^k\,\mathcal{D}^{(\mathcal{V})}_\mu A_k{}^{\beta} - \tfrac{1}{\sqrt{v_1}}\sinh\eta\,\mathrm{i}\,\sigma_{3\,j}{}^k\,A_k{}^{\beta}\Bigr] \, ,
\end{equation}
where the derivative is covariantized with respect to the background $U(1)_{\rm R} \subset$ SU(2)$_\mathrm{R}$ symmetry only. In turn this gives a contribution to the symbol of $D_{10}^\mathrm{hyp}$ of 
\begin{equation}
4\,\Xi_i{}^{\alpha} \bigl(\widehat{Q}\Xi_i{}^{\alpha}\bigr)^\dagger \; \ni \; 8\,\varepsilon^{ij}\Omega_{\alpha\beta}\,\Xi_i{}^{\alpha}\,\check{K}^\mu{}_j{}^k\,\partial_\mu A_k{}^{\beta} \, ,
\end{equation}
coming from the second term in \eqref{eq:hyp-deform}. Evaluating the bilinears $\check{K}^\mu{}_j{}^k$, we obtain the following symbol matrix $\sigma[\,D_{10}^\mathrm{hyp}\,]$,
\begin{equation}
\mathbb{X}_1^{{\rm P}}\,
\begin{pmatrix}
p_1 & \cosh\eta\,p_2 & \cosh\eta\,p_3 & \cosh\eta\,p_4 \\
-\cosh\eta\,p_2 & p_1 & \cosh\eta\,p_4 & -\cosh\eta\,p_3 \\
-\cosh\eta\,p_3 & -\cosh\eta\,p_4 & p_1 & \cosh\eta\,p_2 \\
-\cosh\eta\,p_4 & \cosh\eta\,p_3 & -\cosh\eta\,p_2 & p_1
\end{pmatrix} 
\mathbb{X}_0^{{\rm P}} \, ,
\end{equation}
with the fundamental bosons and fermions arranged as
\begin{equation}
\mathbb{X}^{\rm P}_0 = \begin{pmatrix} A_1{}^{1} - A_2{}^{2} \\ \mathrm{i}\bigl(A_1{}^{1} + A_2{}^{2}\bigr) \\ -\mathrm{i}\bigl(e^{-\mathrm{i}\tau}A_1{}^{2} - e^{\mathrm{i}\tau}A_2{}^{1}\bigr) \\ e^{-\mathrm{i}\tau}A_1{}^{2} + e^{\mathrm{i}\tau}A_2{}^{1} \end{pmatrix} \, , \quad
\mathbb{X}^{\rm P}_1 = \begin{pmatrix} -8\mathrm{i}\bigl(\Xi_1{}^{1} - \Xi_2{}^{2}\bigr) \\ 8\bigl(\Xi_1{}^{1} + \Xi_2{}^{2}\bigr) \\ -8\bigl(e^{-\mathrm{i}\tau}\,\Xi_1{}^{2} - e^{\mathrm{i}\tau}\,\Xi_2{}^{1}\bigr) \\ -8\mathrm{i}\bigl(e^{-\mathrm{i}\tau}\,\Xi_1{}^{2} + e^{\mathrm{i}\tau}\,\Xi_2{}^{1}\bigr) \end{pmatrix}  \, .
\end{equation}
Similar to the vector multiplet calculation in section \ref{sec:index-result}, the above fields are real and neutral under the R-transformation of the $\widehat{Q}^2$-algebra. At the origin $\eta = 0$, the symbol matrix reduces to
\begin{equation}
\begin{pmatrix}
p_1 & p_2 & p_3 & p_4 \\
-p_2 & p_1 & p_4 & -p_3 \\
-p_3 & -p_4 & p_1 & p_2 \\
-p_4 & p_3 & -p_2 & p_1
\end{pmatrix} \, .
\end{equation}
We recognize the symbol of the ASD complex at the fixed point of the $\widehat{Q}^2$ action, which shows that the equivariant index of $D_{10}^\mathrm{hyp}$ is given by the opposite of the index of $D_{10}^\mathrm{vec}$ \eqref{eq:ind-vec-AS}, as in \cite{Pestun:2007rz,Pestun:2016zxk}. Thus, we conclude that a physical hypermultiplet will contribute an inverse factor of \eqref{eq:vec-one-loop-final} to the one-loop determinant,
\begin{equation}\label{eq:finalanswerphysicalhyper}
Z_{\text{1-loop}}^{\rm hyp}(X_+,X_-) = \bigl(v_1\,\chi_{\rm V}(X_+,X_-)\bigr)^{-1/4} \, .
\end{equation}
Note that we could have started by allowing a coupling of the physical hypermultiplets to vector multiplets \cite{deWit:1999fp}, similar to the situation of the previous subsection. Going over the calculation of the compensating hypermultiplet with the reality condition \eqref{eq:sec-real-cond-phys} instead of \eqref{eq:sec-real-cond-comp} we obtain once again \eqref{eq:finalanswerphysicalhyper} above, leading us to conclude that the explicit coupling does not result in a change of the one-loop determinant. This is of course natural to understand from the fact that the one-loop determinant only takes in account quadratic fluctuations around the localization locus and not higher order interactions. The generalization to Riemann surface horizons also follows in complete analogy to section \ref{subsec:genus}.

We close the hypermultiplet analysis by remarking that, both for the compensating and physical hypermultiplets, the Atiyah-Singer index theorem greatly simplifies the computation of the one-loop determinants by allowing us to focus only on the symbol of the relevant differential operators. While an explicit mode analysis as in section \ref{sec:modes} should still be possible in principle, the differential operators involved and the corresponding systems of ODEs on the radial modes look technically more involved due to the proliferation of terms arising in the truncation to quadratic order of the vector multiplet couplings.

\section{The quantum entropy function in gauged supergravity}
\label{sec:final}
\setcounter{equation}{0}

In sections \ref{sec:vectors} and \ref{sec:hypers}, we have obtained the one-loop determinant for an arbitrary number $n_V + 1$ of abelian vector multiplets, the hypermultiplet compensator and an arbitrary number $n_H$ of physical hypermultiplets in the black hole near-horizon background. The result depends on the off-shell fluctuations of the scalar fields of the vector multiplets via the K\"{a}hler potential $\chi_{\rm V}$ of the theory, as explained in section \ref{sec:scale-inv}. In addition, the index theorem computation makes it clear that this dependence is captured by the behavior of the fluctuations at the origin $\eta=0$ of the $\mathbb{H}_2$ factor in the near-horizon geometry. Based on the general arguments above, we also expect that the one-loop determinant for other multiplets in the theory, such as the Weyl multiplet and potential massive Kaluza-Klein multiplets, will take a similar form. In the absence of a direct computation, we will parametrize the contribution of such additional multiplets by a number $a_0$. Hence, we use the following one-loop determinant in the localized QEF \eqref{eq:dmacro-loc-diverge},
\begin{equation}
\label{eq:one-loop-full}
Z_{\text{1-loop}}^{\text{full}}(\phi_+,\phi_-) = \Bigl(v_1\,\chi_\mathrm{V}(\phi_+,\phi_-)\Bigr)^{\frac14(n_V + 1 - n_H) + a_0} \, ,
\end{equation}
with $\phi^I_\pm$ coordinates on the localizing manifold. They correspond to the $2n_V + 2$ real vector multiplet scalar fields $X_\pm^I$, including the off-shell BPS fluctuations at the fixed point $\eta=0$ of the $\widehat{Q}^2$ action around the on-shell background. With this result, the black hole degeneracies computed from the QEF take the form
\begin{align}
\label{eq:dmacro-final}
\begin{split}
\!\!\!\!d_{\rm macro}(p^I,q_I) = \int_{-\infty}^{+\infty}&\,\Bigl(\prod_{I=0}^{n_V}\,\mathrm{d}\phi_+^I\Bigr)\;\delta\Bigl(g\,\xi_I\phi^I_+ - \frac{1}{2\sqrt{v_1}}\Bigr)\,\exp\Bigl[-S_{\rm cl}[\,p^I, q_I, \phi^I_+]\Bigr] \; \times \\
&\!\!\!\!\!\!\!\int\,\Bigl(\prod_{I=0}^{n_V}\,\mathcal{D}\phi_-^I\Bigr)\,\Bigl(v_1\,\chi_\mathrm{V}(\phi_+,\phi_-)\Bigr)^{\frac14(n_V + 1 - n_H) + a_0}\,Z_{\rm measure}(\phi_+, \phi_-) \, .
\end{split}
\end{align}
For $n_H = 0$ and a spherical horizon, it was shown in \cite{Hristov:2018lod} that the two-derivative classical action on the localization locus only depends on $\phi^I_+$ and takes the form
\begin{equation}
S_{\rm cl}[\,p^I, q_I, \phi^I_+] = 8\pi^2\sqrt{v_1}\,\Bigl( p^I F_I^+(\phi_+) + q_I \phi^I_+\Bigr) \, ,
\end{equation}
where $F^\pm(X_\pm)$ are the prepotentials of the theory, and $F^\pm_{IJ\ldots}$ denotes successive derivatives with respect to the argument.
It was further shown in \cite{Hristov:2018lod} that a saddle-point evaluation of the first line in \eqref{eq:dmacro-final}, i.e. neglecting the contribution from the one-loop determinant and the measure, correctly reproduces the Bekenstein-Hawking entropy. Using \eqref{eq:one-loop-full}, we can now address the question of extracting the first corrections to the area-law. To do so, we will analyze how the logarithm of $d_{\rm macro}$ scales with the rank $N$ of the dual field theory, which the holographic dictionary relates to bulk quantities as
\begin{equation}
N^{\alpha} \sim (g^2 G_N)^{-1} \, .
\end{equation} 
Above, the positive power $\alpha$ and the proportionality factor are model-dependent and do not concern us directly in the following discussion.
In the bulk supergravity theory, the electromagnetic charges are quantized according to \cite{Hristov:2018lod,Benini:2016rke}
\begin{equation}
2\,g\,\xi_I\,p^I \in \mathbb{Z} \, , \qquad \frac{4\pi}{2\,g\,\xi_I}\,q_I \in \mathbb{Z} \, , 
\end{equation}
and no sum over $I$ is implied. It was also shown in \cite{Hristov:2018lod} using a comparison with the boundary CFT$_3$ that the combination
\begin{equation}
\label{eq:no-scale-FI}
\widetilde{\xi}_I := \kappa^{-1}\,\xi_I \, , \quad {\rm with} \;\; \kappa^2 = 8\pi\,G_N \,
\end{equation}
 is a dimensionless FI parameter, i.e. a pure number. Therefore, the properly quantized charges independent of the scales set by $g$ and $G_N$ are
\begin{equation}
\label{eq:no-scale-charges}
\mathfrak{p}^I := g\,\kappa\,p^I \, , \quad \mathfrak{q}_I := \frac{4\pi}{2\,g\,\kappa}\,q_I \, .
\end{equation}
They satisfy $2\,\widetilde{\xi}_I\,\mathfrak{p}^I \in \mathbb{Z}$ and $(2\,\widetilde{\xi}_I)^{-1}\,\mathfrak{q}_I \in \mathbb{Z}$, where again no sum over $I$ is implied. As a function of these charges, the classical action reads
\begin{equation}
\label{eq:class-action-charges}
S_{\rm cl}[\,\mathfrak{p}^I, \mathfrak{q}_I, \phi^I_+] = 8\pi^2\sqrt{v_1}\,\Bigl(\frac{1}{g\,\kappa}\,\mathfrak{p}^I F_I^+(\phi_+) + \frac{g\,\kappa}{2\pi}\,\mathfrak{q}_I\,\phi^I_+\Bigr) \, ,
\end{equation}
and the degeneracies \eqref{eq:dmacro-final} take the form of a constrained Laplace transform:
\begin{equation}
\label{eq:dmacro-Laplace}
d_{\rm macro}(\mathfrak{p}^I,\mathfrak{q}_I) = \int_{-\infty}^{+\infty}\,\Bigl(\prod_{I=0}^{n_V}\,\mathrm{d}\phi_+^I\Bigr)\,\delta\Bigl(g\,\kappa\,\widetilde{\xi}_I\phi^I_+ - \frac{1}{2\sqrt{v_1}}\Bigr)\,Z_{\rm macro}(\mathfrak{p}^I,\phi^I_+)\,e^{-4\pi\sqrt{v_1}\,g\,\kappa\,\mathfrak{q}_I\phi^I_+} \, ,
\end{equation}
where
\begin{align}
\label{eq:dmacro-integrand}
\begin{split}
\!\!\!\!\!\!\!Z_{\rm macro}(\mathfrak{p}^I,\phi^I_+) =&\; \exp\Bigl[-\frac{8\pi^2\sqrt{v_1}}{g\,\kappa}\,\mathfrak{p}^I F_I^+(\phi_+)\Bigr] \; \times \\
&\;\; \int\,\Bigl(\prod_{I=0}^{n_V}\,\mathcal{D}\phi_-^I\Bigr)\,\Bigl(v_1\,\chi_\mathrm{V}(\phi_+,\phi_-)\Bigr)^{\frac14(n_V + 1) + a_0}\,Z_{\rm measure}(\phi_+,\phi_-) \, .
\end{split}
\end{align}

We will analyze the behavior of \eqref{eq:dmacro-Laplace} when the rank $N$ of the boundary field theory is very large and the boundary charges $(\mathfrak{p}^I,\mathfrak{q}_I)$ are kept fixed. Observe that this limit corresponds to scaling the original bulk charges according to 
\begin{equation}
\label{eq:charge-limit}
(p^I,q_I) \rightarrow (\Lambda\,p^I,\Lambda^{-1}q_I) \, , \quad \textnormal{with} \quad \Lambda := (g\,\kappa)^{-1} \gg 1 \, ,
\end{equation}
while keeping the scales $v_1$ and $v_2$ fixed. By definition of the QEF, we expect that the dominant contribution to \eqref{eq:dmacro-Laplace} in this limit comes from the classical on-shell field configuration \cite{Sen:2008vm}, which is located at $\phi_\pm^I = 2\mathring{X}_\pm^I$ \cite{Hristov:2018lod}. 

According to \eqref{eq:1/2-BPS-vec}, the attractor values of the scalar fields scale as $\mathring{X}_\pm^I \sim \Lambda\,v_1^{-1/2}\,$ in the large $\Lambda$ limit. Due to the homogeneity of the K\"{a}hler potential $\chi_{\rm V}$, the one-loop determinant factor in \eqref{eq:dmacro-integrand} therefore scales as
\begin{equation}
\label{eq:one-loop-scale}
\Bigl(v_1\,\chi_\mathrm{V}(\mathring{X}_+,\mathring{X}_-)\Bigr)^{\frac14(n_V + 1) + a_0} \sim \bigl(\Lambda^2\bigr)^{\frac14(n_V + 1) + a_0} \, .
\end{equation}
Lacking an explicit expression for the measure factor as a function of the $\phi^I_\pm$ coordinates, we will assume for the moment that $Z_{\rm measure} \sim \Lambda^0$ in the large $\Lambda$ limit. The justification we can offer at this stage is to recall that in ungauged supergravity the measure on the localization manifold entering the localized QEF was investigated in more details for a certain class of black holes and found to be $\Lambda$-independent in the corresponding large charge limit \cite{Dabholkar:2011ec,Murthy:2015yfa}. We will comment further on this below.
For large $\Lambda$, the exponential factor in \eqref{eq:dmacro-integrand} scales as $\exp\bigl(\Lambda^2\bigr)$. More precisely, it was shown in \cite{Hristov:2018lod} that we can combine this factor with the exponential factor in \eqref{eq:dmacro-Laplace} and evaluate them together explicitly at the attractor point. Since the magnetic charges are sensitive to the size of the black hole horizon $A_H$, this yields
\begin{equation}
\label{eq:BH-dom}
\exp\Bigl[-8\pi^2\sqrt{v_1}\,\Bigl(\frac{1}{g\,\kappa}\,\mathfrak{p}^I F_I^+(\phi_+) + \frac{g\,\kappa}{2\pi}\,\mathfrak{q}_I\,\phi^I_+\Bigr)\Bigr]_{\phi^I_+ = 2\mathring{X}^I_+} = \exp\Bigl[\frac{A_H}{4\,G_N}\Bigr] \, .
\end{equation}
We also expect a Hessian correction to the leading behavior \eqref{eq:BH-dom} coming from the determinant of the second derivative of the classical action \eqref{eq:class-action-charges} evaluated at the attractor point. The delta-function constraint in \eqref{eq:dmacro-Laplace} reduces the rank of the Hessian matrix from $n_V + 1$ to $n_V$, and we arrive at the following leading contribution to the microscopic degeneracies in the large $\Lambda$ limit:
\begin{equation}
\label{eq:dmacro-saddle}
d_{\rm macro}(\mathfrak{p}^I,\mathfrak{q}_I) \; \sim \; \exp\Bigl[\frac{A_H}{4\,G_N}\Bigr]\;\Bigl(\frac{A_H}{4\,G_N}\Bigr)^{-\frac12 n_V}\,\Bigl(\frac{1}{g^2\,G_N}\Bigr)^{\frac14(n_V + 1) + a_0} \, .
\end{equation}
Observe that the three terms above have different origins. The first is the exponential of the Bekenstein-Hawking entropy, which is the leading contribution in the classical limit. The second is the Hessian correction to the classical saddle-point, and as such also depends on the area of the black hole horizon. The third is the contribution \eqref{eq:one-loop-scale} coming from the one-loop determinants we have computed in this paper, written directly in terms of the relevant bulk quantities $g$ and $G_N$.
We therefore arrive at one of our main results, which is the following expression for the entropy of the asymptotically AdS$_4$ black holes under consideration in the large charge limit \eqref{eq:charge-limit}:
\begin{equation}
\label{eq:result-S2-nohyp}
\log d_{\rm macro}(\mathfrak{p}^I,\mathfrak{q}_I) = \frac{A_H}{4\,G_N} - \frac12 n_V \log\Bigl(\frac{A_H}{4\,G_N}\Bigr) + \Bigl(\frac14(n_V + 1) + a_0\Bigr)\log\Bigl(\frac{1}{g^2\,G_N}\Bigr) + \ldots \, .
\end{equation}
In general, the gauge coupling sets the length scale $L$ of the asymptotic AdS$_4$ spacetime as $L \propto g^{-1}$ \cite{Cacciatori:2009iz}. In turn, the scales $v_1$ and $v_2$ in the near-horizon geometry are determined in terms of $g$ (or $L$) in a model-dependent fashion. We should now recall that in \eqref{eq:result-S2-nohyp}, the third term comes from the scaling behavior of the scalar field attractor values. As is clear from \eqref{eq:1/2-BPS-vec}, the attractor values are only sensitive to $v_1$ and so we are able to identify the scale set by $g$ in this term as the scale of the near-horizon $\mathbb{H}_2$ factor. \\

The lessons from \cite{Hristov:2018lod} and the present paper also allow us to write a generalization of the result \eqref{eq:result-S2-nohyp} to the case where an arbitrary number $n_H$ of physical hypermultiplets are present in the near-horizon background, and when the black hole horizon has the more general topology of a genus-${\rm g}$ Riemann surface $\Sigma_{\rm g}$. We expect the BPS conditions on the former to produce $n_H$ additional delta-function constraints in the localized QEF \eqref{eq:dmacro-final}, further reducing the rank of the Hessian matrix governing the first correction term in \eqref{eq:result-S2-nohyp}. The topology of the horizon affects the one-loop determinant in the way already discussed in section \ref{subsec:genus}. Therefore, in the more general case, we expect the following log-corrections to the Bekenstein-Hawking entropy of asymptotically AdS$_4$ black holes with AdS$_2 \times \Sigma_{\rm g}$ near-horizon geometry:
\begin{eqnarray}
\label{eq:result-gen}
\begin{split}
\log d_{\rm macro} (\mathfrak{p}^I,\mathfrak{q}_I,{\rm g}) =&\; \frac{A_{H,{\rm g}}}{4\,G_N} - \frac12 (n_V-n_H) \log\Bigl(\frac{A_{H,{\rm g}}}{4\,G_N}\Bigr) \\ 
&\; +  (1-{\rm g})\Bigl(\frac14(n_V + 1 - n_H)  + a_0\Bigr)\log\Bigl(\frac{1}{g^2\,G_N}\Bigr) + \ldots \, .
\end{split}
\end{eqnarray}
Above we have also extracted a factor of $(1-{\rm g})$ from the unknown coefficient $a_0$. This is expected from the arguments based on the Atiyah-Singer index theorem in section \ref{subsec:genus}: the one-loop determinant of the Weyl and additional Kaluza-Klein multiplets will be sensitive to the topology of the horizon through its Euler characteristic.\\ 

We close this section with a number of comments. The first is that if we relax the assumption that the measure on the localizing manifold is $\mathcal{O}(1)$ and instead assume a scaling exponent $a_{\rm m}$ in the large $\Lambda$ limit, we would pick up an additional term in the coefficient of the $\log(g^2 G_N)$ term in \eqref{eq:result-gen}. However for the measure we have no strong reason to expect that $a_{\rm m}$ would come multiplied by a factor of $(1-{\rm g})$ for a horizon $\Sigma_{\rm g}$. This is because, while the one-loop determinants are sensitive to the topology of the black hole horizon as shown in this paper, the measure on the localizing manifold is a property of the off-shell BPS field configuration.\footnote{Explicit expressions for the measure factor in ungauged supergravity have been put forward in \cite{Dabholkar:2011ec,Murthy:2015yfa}, see also \cite{LopesCardoso:2006ugz,Cardoso:2008fr} for a more general discussion based on duality invariance of black hole partition functions.} Since the latter only depends on the genus through the linear constraint satisfied by the magnetic charges \cite{Caldarelli:1998hg,Benini:2016rke}, it is not a priori clear how a non-trivial scaling coefficient will be sensitive to the genus of the Riemann surface $\Sigma_{\rm g}$. We view this as another justification for setting $a_{\rm m}$ to zero when extracting the log corrections, as discussed below \eqref{eq:one-loop-scale}. Evidently a more thorough analysis of the measure is desirable to properly understand these aspects, and we plan on investigating this in the future.

Logarithmic corrections to the Bekenstein-Hawking entropy of certain asymptotically AdS$_4$ solutions have been computed, both in 11d supergravity using zero-mode counting \cite{Jeon:2017aif,Liu:2017vbl} and in the dual field theory at the level of the topologically twisted index \cite{Liu:2017vll,Liu:2018bac,PandoZayas:2019hdb,Hosseini:2018qsx}.  
Comparison with previous results for specific models would first require us to fix the value of the $a_0$ coefficient in \eqref{eq:result-gen}. The Weyl multiplet will bring a universal (model-independent) contribution to $a_0$, and we expect that it can be computed using the methods put forward in \cite{Jeon:2018kec,deWit:2018dix} and adapted to gauged supergravity. Each Kaluza-Klein multiplet resulting from the embedding will also affect the value of $a_0$, and the total contribution will be model-dependent. Once these contributions are known, it will be most interesting to see how \eqref{eq:result-gen} compares to the results mentioned above.

Another interesting avenue to explore is the case of refinement with angular momentum. Unlike the asymptotically flat black holes, solutions in gauged supergravity admit refinement with angular momentum while still preserving the same number of supercharges \cite{Hristov:2018spe}. This allows us to use the method of supersymmetric localization on these more general solutions, and indeed we already made use of the refinement in section \ref{sec:FP-result} as one way of evaluating the one-loop determinants. To evaluate the full quantum entropy function with refinement one however needs to start again from the classical action and repeat the steps in \cite{Hristov:2018lod}. In this case it would also be interesting to compare the off-shell analysis with the compelling new evidence that the on-shell supergravity action depends entirely on the underlying topology of the solution and the fixed points of the supersymmetric Killing vector \cite{BenettiGenolini:2019jdz}. 

\section*{Acknowledgements}

We would like to thank N. Bobev, B. de Wit, S.M. Hosseini, N. Mekareeya, S. Murthy  and A. Zaffaroni for useful and stimulating discussions. KH is supported in part by the Bulgarian NSF grants DN08/3 and N28/5. VR is supported in part by INFN and by the ERC Starting Grant 637844-HBQFTNCER.

\newpage

\appendix

\section{Black hole solution and Killing spinors}\label{app:A}
\setcounter{equation}{0}

Here we summarize the Euclidean near-horizon solution in the conformal supergravity formalism, as given in \cite{Hristov:2018lod}. This on-shell field configuration pertains to the bosonic fields of the Weyl multiplet (comprising the metric and various gauge and auxiliary fields in the superconformal setup), and to the bosonic fields of vector multiplets and hypermultiplets coupled to the conformal supergravity background.

The near-horizon geometry is $\mathbb{H}_2\times S^2$, and the metric can be written in hyperbolic disk coordinates,
\begin{equation}
\label{eq:1/2-BPS-metric}
ds^2 = v_1\,\bigl(\sinh^2\eta\,\mathrm{d}\tau^2 + \mathrm{d}\eta^2\bigr) + v_2\,\bigl(\mathrm{d}\theta^2 + \sin^2\theta\,\mathrm{d}\varphi^2\bigr) \, ,
\end{equation}
with $v_1$ and $v_2$ are real positive constants parameterizing the sizes of the $\mathbb{H}_2$ and $S^2$ spaces, respectively. In this coordinate system, the black hole horizon sits at $\eta=0$. We use the vielbein one-forms
\begin{equation}
e^1 = \sqrt{v_1}\sinh\eta\,\mathrm{d}\tau \, , \quad e^2 = \sqrt{v_1}\,\mathrm{d}\eta \, , \quad e^3 = \sqrt{v_2}\,\mathrm{d}\theta \, , \quad e^4 = \sqrt{v_2}\sin\theta\,\mathrm{d}\varphi \, .
\end{equation}
After a gauge choice for the dilatation transformations (see \cite{Hristov:2018lod}), we take the non-vanishing components of the auxiliary tensor field $T_{ab}$ in the Weyl multiplet to be
\begin{equation}
\label{eq:1/2-BPS-T-tensor}
T_{12}^\mp = \pm\frac{2}{\sqrt{v_1}} \, , \quad T_{34}^\mp = -\frac{2}{\sqrt{v_1}} \, ,
\end{equation}
in which $\pm$ correspond to (anti-)self-dual projections. To gauge-fix the special conformal boost symmetry we set the gauge field for dilatations to zero, and to gauge-fix the SO(1,1)$_\mathrm{R}$ symmetry we set the corresponding gauge field to zero.
The auxiliary scalar $D$ is given by \cite{deWit:2011gk}
\begin{equation}
\label{eq:1/2-BPS-D}
D = -\frac16\,\bigl(v_1^{-1} + 2\,v_2^{-1}\bigr) \, .
\end{equation}
In gauged conformal supergravity, the SU(2)$_\mathrm{R}$ gauge field is expressed in terms of the background gauge fields in the vector multiplets $\mathring{W}_\mu^I$ (see \eqref{eq:1/2-BPS-vec} below) as \cite{deWit:2011gk}
\begin{equation}
\label{eq:1/2-BPS-SU(2)-connection}
\mathcal{V}_\mu{}^i{}_j = -2\mathrm{i}\,g\,\xi_I\mathring{W}_\mu^I\,\sigma_3{}^i{}_j \, ,
\end{equation}
where $\xi_I$ are the Fayet-Illiopoulos parameters and $g$ is the gauge coupling. 

It was shown in \cite{deWit:2011gk, Hristov:2018lod} that the above bosonic field configuration is half-BPS. Among the conformal Killing spinors associated to this geometry, we pick a particular one that we denote $(\xi^i,\,\kappa^i)$ (where $i=1,2$ is the SU(2)$_\mathrm{R}$ symmetry index) to parameterize the localizing supercharge $Q$ used in the main text. It is given by \cite{Hristov:2018lod}
\begin{equation}
\label{eq:loc-CKS}
\xi^1 = \begin{pmatrix} e^{-\tfrac12\mathrm{i}\tau}\cosh\bigl(\frac{\eta}{2}\bigr) \\ -\mathrm{i}\,e^{-\tfrac12\mathrm{i}\tau}\sinh\bigl(\frac{\eta}{2}\bigr) \\ 0 \\ 0 \end{pmatrix} \, , \qquad \xi^2 = \begin{pmatrix} 0 \\ 0 \\ -e^{\tfrac12\mathrm{i}\tau}\sinh\bigl(\frac{\eta}{2}\bigr) \\ -\mathrm{i}\,e^{\tfrac12\mathrm{i}\tau}\cosh\bigl(\frac{\eta}{2}\bigr) \end{pmatrix} \, , \qquad \kappa^i = -\frac{\mathrm{i}}{2}\,\Slash{\mathcal{D}}\xi^i \, .
\end{equation}
The derivative $\mathcal{D}_\mu$ is covariantized with respect to Lorentz and SU(2)$_\mathrm{R}$ transformations,
\begin{equation}
 \mathcal{D}_\mu\xi^i = \partial_\mu\xi^i - \frac14\,\omega_\mu{}^{ab}\,\gamma_{ab}\,\xi^i + \frac12\,\mathcal{V}_\mu{}^i{}_j\,\xi^j \, .
 \end{equation}

We use one hypermultiplet compensator to ensure that the superconformal theory is gauge-equivalent to Poincar\'{e} supergravity. The scalar sections $A_i{}^\alpha$ of this hypermultiplet can be taken constant on the half-BPS background by an SU(2)$_\mathrm{R}$ gauge choice,
\begin{equation}
\label{eq:1/2-BPS-hyp}
\chi_\mathrm{H}^{-1/2} \mathring{A}_i{}^\alpha = \delta_i{}^\alpha \, ,
\end{equation}
where the hyper-K\"{a}hler potential is defined as $\chi_\mathrm{H} := \tfrac12\,\varepsilon^{ij}\,\Omega_{\alpha\beta}\,\mathring{A}_i{}^\alpha\,\mathring{A}_j{}^\beta$. This choice breaks the SU(2)$_{\rm R}$ invariance of the background down to $U(1)_{\mathrm{R}}$.

We also consider $n_V + 1$ abelian vector multiplets, including the conformal compensator. The half-BPS bosonic field configuration satisfies \cite{Hristov:2016vbm}
\begin{equation}
\label{eq:1/2-BPS-vec}
g\,\xi_I\mathring{F}_{34}^{\mp\,I} = \frac{1}{4 v_2} \, , \qquad g\,\xi_I\mathring{X}^I_\mp = \frac{1}{4\sqrt{v_1}} \, ,
\end{equation}
with $\mathring{F}_{\mu\nu}^I$ field strengths of the vector fields $\mathring{W}_\mu^I$. In the Euclidean superconformal formalism, the coupling of vector multiplets to the gravity background is specified by two homogeneous functions each of degree two in the scalar fields $X_+^I$ and $X_-^I$ \cite{deWit:2017cle}. At the two derivative level, such prepotentials $F^\pm(X_\pm)$ completely determine the action of the theory under consideration. The associated K\"{a}hler potential has Weyl weight two and reads
\begin{equation}
\chi_{\rm V}(X_+,X_-) := X^I_+\,F_I^-(X_-) + X^I_-\,F_I^+(X_+) \, ,
\end{equation}
where $F^\pm_{IJ\ldots}$ denotes successive derivatives with respect to the argument.

\section{Kernel and cokernel of $D_{10}$ for the vector multiplet}
\label{app:B}
\setcounter{equation}{0}

Here we explicitly solve the kernel and cokernel equations derived in sections \ref{sec:kerneq} and \ref{sec:cokerneq}. Using the mode expansions discussed in \ref{sec:mod-exp-bc}, we reduce the problem to a system of ordinary differential equations along the radial coordinate $\eta$ of the $\mathbb{H}_2 \times S^2$ geometry. Imposing boundary and smoothness conditions for the solutions of the coupled ODEs, we obtain the dimensions of the kernel and cokernel of $D_{10}^\mathrm{vec}$.

\subsection{Solving the kernel equations}
\label{app:kernel-modes}

We begin with the kernel equation associated with the ghost field $c$ \eqref{eq:kerneq-c}. It takes the form of a ``Laplacian'' operator $\nabla_\mu \frac{1}{K^2} \nabla^\mu$ acting on the scalar field
\begin{equation}
R := \mathring{v}^\mu\tilde{W}_\mu - 4\,\sigma \, .
\end{equation}
Both $\sigma$ and $\tilde{W}_\tau$ are scalars on the $S^2$ factor of the near-horizon geometry and are neutral under SU(2)$_\mathrm{R}$, so we use the mode expansion \eqref{eq:scal-mod-exp} to write
\begin{equation}
R = R(\eta)\,e^{\mathrm{i}n\tau}\,Y_\ell{}^m(\theta,\varphi) \, , \quad R(\eta) := \tfrac{2}{\sqrt{v_1}}\,\tilde{W}_\tau(\eta) - 4\,\sigma(\eta) \, .
\end{equation}
Our Laplacian operator then acts on $R$ according to
\begin{align}
\label{eq:our-scal-lap}
\nabla_\mu&\Bigl[\frac{1}{K^2}\,\partial^\mu R\Bigr] = \\
&\frac{1}{K^2 v_1}\Bigl[R(\eta)'' + \bigl(\coth(\eta) - 2\tanh(\eta)\bigr) R(\eta)' - \Bigl(\frac{n^2}{\sinh^2\eta} + \frac{v_1}{v_2}\,\ell(\ell+1)\Bigr) R(\eta)\,\Bigr]e^{\mathrm{i}n\tau}\,Y_\ell{}^m \, , \nonumber 
\end{align}
where here and below a prime denotes a derivative with respect to $\eta$, and we have used the eigenvalue equation on $S^2$ of radius $v_2$,
\begin{equation}
-\nabla_{S^2}^2 Y_\ell{}^m(\theta,\varphi) = \frac{\ell(\ell+1)}{v_2}\,Y_\ell{}^m(\theta,\varphi) \, .
\end{equation}
We will denote the differential operator in the bracket of \eqref{eq:our-scal-lap} by $\Delta^{(n,\ell)}$ so that we can write \eqref{eq:kerneq-c} as a differential equation on each radial mode
\begin{equation}
\label{eq:delta-nl}
\Delta^{(n,\ell)}R(\eta) = 0 \, .
\end{equation}
This equation can be solved explicitly and the result is given in terms of the hypergeometric function ${}_2F_1$ which governs the eigenfunctions of the scalar Laplacian on $\mathbb{H}_2$ (see e.g. \cite{Banerjee:2011jp}), on which we then need to impose the boundary and smoothness conditions of section \ref{sec:mod-exp-bc}. But since we are only interested in the number of solutions for a given pair $(n,\ell)$ and not their explicit forms, we can use the analysis of \cite{David:2016onq} to show that there are no solutions for $(n,\ell) \neq (0,0)$ and only a constant solution for $R(\eta)$ when $(n,\ell) = (0,0)$. Explicitly, we can multiply \eqref{eq:delta-nl} by $-R(\eta)\tanh\eta\,\cosh^{-1}\eta$ to write the left-hand side as
\begin{equation}
-\partial_\eta\Bigl[\frac{\tanh\eta}{\cosh\eta}\,R(\eta)\,R(\eta)'\Bigr] + \frac{\tanh\eta}{\cosh\eta}\,\Bigl(\bigl(R(\eta)'\bigr)^2 + \Bigl(\frac{n^2}{\sinh^2\eta} + \frac{v_1}{v_2}\,\ell(\ell+1)\Bigr) R(\eta)^2\Bigr) \, .
\end{equation}
When $R(\eta)$ satisfies \eqref{eq:delta-nl}, this quantity must vanish. Integrating over $\eta$, we then have
\begin{equation}
\int_0^{\infty}\frac{\tanh\eta}{\cosh\eta}\,\Bigl(\bigl(R(\eta)'\bigr)^2 + \Bigl(\frac{n^2}{\sinh^2\eta} + \frac{v_1}{v_2}\,\ell(\ell+1)\Bigr) R(\eta)^2\Bigr)\,\mathrm{d}\eta - \Bigl[\frac{\tanh\eta}{\cosh\eta}\,R(\eta) \,R(\eta)'\Bigr]_0^{\infty} = 0 \, .
\end{equation}
With the boundary conditions \eqref{eq:norm-bc-vec-bos} and the smoothness conditions \eqref{eq:smooth-vec-bos} we see that the boundary term vanishes. The remaining integral is positive-definite, so for $(n,\ell) \neq (0,0)$ we must have $R(\eta) = 0$. When $(n,\ell) = (0,0)$, it is sufficient to have $R(\eta)' = 0$ and so a constant solution is allowed in this case. Since $\ell = 0$ implies $m= 0$ for the spherical harmonics and since $Y_0{}^0(\theta,\varphi)$ is a constant, we conclude that \eqref{eq:kerneq-c} admits a single constant solution for the scalar combination
\begin{equation}
\label{eq:sol-kerneq-c}
\mathring{v}^\mu\tilde{W}_\mu - 4\,\sigma = \begin{cases} C_1 \quad &\textnormal{for} \;\; (n,\ell,m) = (0,0,0)  \\ 0  \quad &\textnormal{for} \;\; (n,\ell,m) \neq (0,0,0)  \end{cases}\;.
\end{equation}

We now turn to \eqref{eq:kerneq-lambda}. Evaluating explicitly the $K^{\mu\nu}_{ij}$ bilinear, we obtain a set of three equations (as expected from an SU(2) triplet):
\begin{align}
\begin{split}
\label{eq:kerneq-lambda-1}
\tilde{F}_{12} + \cosh\eta\,\tilde{F}_{34} - 2\,\sinh\eta\,\partial_2\sigma =&\; 0 \, , \\[1mm]
\tilde{F}_{13} - \cosh\eta\,\tilde{F}_{24} - 2\,\sinh\eta\,\partial_3\sigma =&\; 0 \, , \\[1mm]
\tilde{F}_{14} + \cosh\eta\,\tilde{F}_{23} - 2\,\sinh\eta\,\partial_4\sigma =&\; 0 \, ,
\end{split}
\end{align}
and all indices are in tangent space. We can make use of the previous result \eqref{eq:sol-kerneq-c} to analyze these equations as follows. Since $\partial_\mu(\mathring{v}^\nu\tilde{W}_\nu) = 4\,\partial_\mu\sigma$ for any $(n,\ell,m)$, we may trade the derivatives on $\tilde{W}_\tau$ for derivatives on the scalar field $\sigma$. In addition, $\tilde{W}_\eta$ is a scalar on $S^2$ so we can use the expansion along spherical harmonics \eqref{eq:scal-mod-exp},
\begin{equation}
\tilde{W}_\eta = \tilde{W}_\eta(\eta)\,e^{\mathrm{i}n\tau}\,Y_\ell{}^m(\theta,\varphi) \, .
\end{equation}
On the other hand, the fields $\tilde{W}_\theta$ and $\tilde{W}_\varphi$ form a vector on the 2-sphere so we expand them in the basis \eqref{eq:vec-mod-exp},
\begin{align}
\tilde{W}_\theta =&\; W_B(\eta)\,e^{\mathrm{i}n\tau}\,\partial_\theta Y_\ell{}^m + W_C(\eta)\,e^{\mathrm{i}n\tau}\,\frac{1}{\sin\theta}\,\partial_\varphi Y_\ell{}^m \, , \nonumber \\
\tilde{W}_\varphi =&\; W_B(\eta)\,e^{\mathrm{i}n\tau}\,\partial_\varphi Y_\ell{}^m - W_C(\eta)\,e^{\mathrm{i}n\tau}\,\sin\theta\,\partial_\theta Y_\ell{}^m \, .
\end{align}
This leads to a mode expansion for the various field strengths entering \eqref{eq:kerneq-lambda-1}. Having eliminated $\tilde{W}_\tau$ using \eqref{eq:sol-kerneq-c}, we obtain the following system of ODEs on the radial modes:
\begin{align}
0 =&\; \frac{\mathrm{i}\,n}{\sinh\eta\cosh\eta}\,\tilde{W}_\eta(\eta) - 2\sqrt{v_1}\,\coth\eta\;\sigma(\eta)' + \frac{v_1}{v_2}\,\ell(\ell+1)\,W_C(\eta) \, , \nonumber \\
0 =&\; \frac{\mathrm{i}\,n}{\sinh\eta\cosh\eta}\,W_B(\eta) + W_C(\eta)' - 2\sqrt{v_1}\,\coth\eta\;\sigma(\eta) \, , \\
0 =&\; \frac{\mathrm{i}\,n}{\sinh\eta\cosh\eta}\,W_C(\eta) - W_B(\eta)' + \tilde{W}_\eta(\eta) \, . \nonumber
\end{align}

The last kernel equation to analyze is the gauge-fixing \eqref{eq:kerneq-b}. Using the mode decomposition and \eqref{eq:sol-kerneq-c}, and after some straightforward manipulations, it leads to the radial ODE:
\begin{equation}
0 = \frac{2\mathrm{i}\,n\,\sqrt{v_1}}{\sinh^2\eta}\,\sigma(\eta) + \tilde{W}_\eta(\eta)' + \bigl(\coth\eta - 2\tanh\eta\bigr)\,\tilde{W}_\eta(\eta) - \frac{v_1}{v_2}\,\ell(\ell+1)\,W_B(\eta) \, . \vspace{3mm}
\end{equation}

To summarize, by making use of \eqref{eq:sol-kerneq-c}, we have shown that the kernel equations \eqref{eq:kerneq-b} and \eqref{eq:kerneq-lambda} are equivalent to a system of coupled first-order ODEs for the radial modes,
\begin{equation}
\label{eq:diff-eq-sys}
\vec{u}(\eta)' = A_{n,\ell}(\eta) \cdot \vec{u}(\eta) \, , 
\end{equation}
with
\begin{equation}
\vec{u}(\eta) = \Bigl(2\sqrt{v_1}\,\sigma(\eta) \; , \; \tilde{W}_\eta(\eta) \; , \; W_B(\eta) \; , \; W_C(\eta)\Bigr)^\mathrm{T} \, , 
\end{equation}
and
\begin{equation}\label{eq:matrixA}
A_{n,\ell}(\eta) = \begin{pmatrix} 0 & \frac{\mathrm{i}n}{\cosh^2\eta} & 0 & \frac{v_1}{v_2}\ell(\ell+1)\tanh\eta \\ -\frac{\mathrm{i}n}{\sinh^2\eta} & 2\tanh\eta - \coth\eta & \frac{v_1}{v_2}\ell(\ell+1) & 0 \\ 0 & 1 & 0 & \frac{\mathrm{i}n}{\sinh\eta\cosh\eta} \\ \coth\eta & 0 & -\frac{\mathrm{i}n}{\sinh\eta\cosh\eta} & 0 \end{pmatrix} \, . \vspace{2mm}
\end{equation}
We will now discuss potential solutions to this system when imposing the boundary and smoothness conditions of Section \ref{sec:mod-exp-bc}. The discussion splits depending on the values of the quantum numbers $(n,\ell)$.\\

\underline{\textbf{The case} $\mathbf{(n,\ell)=(0,0)}$}:\\

In this case, \eqref{eq:diff-eq-sys} simplifies and admits the following solutions: 
\begin{align}
\sigma(\eta) =&\; C_2 \, , \qquad \tilde{W}_\eta(\eta) = C_3\cosh\eta\coth\eta \, , \nonumber \\
W_B(\eta) =&\; C_4 + C_3\Bigl(\cosh\eta + \log\tanh\frac{\eta}{2}\Bigr) \, , \quad W_C(\eta) =  C_5 + 2\sqrt{v_1}\,C_2\log\sinh\eta \, ,
\end{align}
where the $C_i$ are arbitrary real constants. We now summon the boundary conditions \eqref{eq:norm-bc-vec-bos}, which effectively force all the integration constants to vanish. So we are left with the trivial solution,
\begin{equation}
\sigma(\eta) = \tilde{W}_\eta(\eta) = \tilde{W}_\theta(\eta) = \tilde{W}_\varphi(\eta) = 0 \, .
\end{equation}
We can now use \eqref{eq:sol-kerneq-c} to obtain that the last remaining field $\tilde{W}_\tau$ is constant, proportional to $C_1$. This however violates the smoothness condition $\tilde{W}_\tau(\eta) \sim \eta^2$ when $\eta \rightarrow 0$ unless $C_1 =0$.\\ 

In conclusion, when $(n,\ell) = (0,0)$, we have found that there are no non-trivial solutions satisfying the boundary and smoothness conditions in the kernel of $D^\mathrm{vec}_{10}$.\\

\underline{\textbf{The case} $\mathbf{n=0,\, \ell \neq 0}$}:\\

In this case the systems of ODEs on $\{\sigma(\eta),W_C(\eta)\}$ and $\{\tilde{W}_\eta(\eta),W_B(\eta)\}$ decouple. The latter reads:
\begin{equation}
\tilde{W}_\eta(\eta) = W_B(\eta)' \, , \quad \tilde{W}_\eta(\eta)' = \bigl(2\tanh\eta - \coth\eta\bigr)\,\tilde{W}_\eta(\eta) + \frac{v_1}{v_2}\,\ell(\ell+1)\,W_B(\eta) \, ,
\end{equation}
which is equivalent to
\begin{equation}
\tilde{W}_\eta(\eta) = W_B(\eta)' \, , \qquad \Delta^{(0,\ell)}W_B(\eta) = 0 \, .
\end{equation}
We already showed that when $\ell \neq 0$, the only solution to the above equations compatible with the boundary conditions is $W_B(\eta) = 0$ and $\tilde{W}_\eta(\eta) = 0$. On the other hand, the system on $\{\sigma(\eta),W_C(\eta)\}$  can be written as
\begin{equation}
2\sqrt{v_1}\,\sigma(\eta) = \tanh\eta\;W_C(\eta)' \, , \quad W_C(\eta)'' + \frac{1}{\sinh\eta\cosh\eta}\,W_C(\eta)' - \frac{v_1}{v_2}\,\ell(\ell+1)\,W_C(\eta) = 0 \, .
\end{equation}
Multiplying the equation on $W_C(\eta)$ by $-\tanh\eta\,W_C(\eta)$, we obtain
\begin{equation}
-\partial_\eta\Bigl[\tanh\eta\,W_C(\eta)\,W_C(\eta)'\Bigr] + \tanh\eta\,\Bigl(\bigl(W_C(\eta)'\bigr)^2 + \frac{v_1}{v_2}\,\ell(\ell+1)\,\bigl(W_C(\eta)\bigr)^2\Bigr) = 0 \, ,
\end{equation}
which, upon integrating over the radial coordinate, yields a positive-definite integral since the boundary term drops out owing to our boundary conditions. This integral can only vanish for $W_C(\eta) = 0$, which in turn implies $\sigma(\eta) = 0$. Finally, \eqref{eq:sol-kerneq-c} and smoothness of $\tilde{W}_\tau(\eta)$ shows that we must have $\tilde{W}_\tau(\eta) = 0$. In summary, when $n=0$ and $\ell \neq 0$, we only have the trivial solution
\begin{equation}
\sigma(\eta) = \tilde{W}_\tau(\eta) = \tilde{W}_\eta(\eta) = \tilde{W}_\theta(\eta) = \tilde{W}_\varphi(\eta) = 0 \, ,
\end{equation}
in the kernel of $D_{10}^\mathrm{vec}$. \\

\underline{\textbf{The case} $\mathbf{n \neq 0,\, \ell=0}$}:\\

This is another decoupled case. We have a system on $\{\sigma(\eta),\tilde{W}_\eta(\eta)\}$ which can be written as
\begin{equation}
\label{eq:sys-n0}
\tilde{W}_\eta(\eta) = \frac{\cosh^2\eta}{\mathrm{i} n}\,2\sqrt{v_1}\,\sigma(\eta)' \, , \quad \sigma(\eta)'' + \coth\eta\;\sigma(\eta)' - \frac{n^2}{\cosh^2\eta\sinh^2\eta}\sigma(\eta) = 0 \, .
\end{equation}
The differential equation on $\sigma(\eta)$ can be solved explicitly, and the solution is
\begin{equation}
\sigma(\eta) = C_6\tanh^{|n|}\eta\;{}_2F_1\Bigl(\frac12|n| + \frac14\bigl(1 - \sqrt{1 + 4n^2}\bigr),\frac12|n| + \frac14\bigl(1 + \sqrt{1 + 4n^2}\bigr);1+|n|;\tanh^2\eta\Bigr) \, ,
\end{equation}
where $C_6$ is the integration constant and ${}_2F_1$ is the standard hypergeometric function. However, this solution goes to a constant for $\eta \rightarrow \infty$, so compatibility with our choice of boundary conditions \eqref{eq:norm-bc-vec-bos} demands $C_6 = 0$. This implies $\sigma(\eta) = \tilde{W}_\eta(\eta) = 0$. Then we are left with a simple system on $\{W_B(\eta),W_C(\eta)\}$,
\begin{equation}
W_B(\eta)' = \frac{\mathrm{i}n}{\sinh\eta\cosh\eta}\,W_C(\eta) \, , \quad W_C(\eta)' = -\frac{\mathrm{i}n}{\sinh\eta\cosh\eta}\,W_B(\eta) \, ,
\end{equation}
which can be solved explicitly. One finds that the solutions also go to a constant when $\eta \rightarrow \infty$, thus once again our boundary conditions force $W_B(\eta)$ and $W_C(\eta)$ to vanish. In conclusion, for $n \neq 0$ and $\ell = 0$, we only have with the trivial solution
\begin{equation}
\sigma(\eta) = \tilde{W}_\tau(\eta) = \tilde{W}_\eta(\eta) = \tilde{W}_\theta(\eta) = \tilde{W}_\varphi(\eta) = 0 \, ,
\end{equation}
in the kernel of $D_{10}^\mathrm{vec}$.\\

\underline{\textbf{The case} $\mathbf{n \neq 0,\, \ell \neq 0}$}:\\

In the most general case, it is not easy to find an explicit solution of the system \eqref{eq:diff-eq-sys}. A numerical analysis (which we will not present here) hints at the absence of any non-trivial solutions satisfying the boundary conditions. To confirm this result analytically, we can make an asymptotic analysis of the differential system. Specifically, we can use the corollary VII-3-7 of \cite{Hsieh-Sibuya}. Since the matrix \eqref{eq:matrixA} is continuous for any non-zero value of $\eta$ and its limit for $\eta\to\infty$ denoted by $A^\infty_{\ell}$ exists, the hypotheses of the corollary are satisfied. Then, for every non-trivial solution $\vec{u}_*(\eta)$ of the system, one has:
\begin{equation}
\lim_{\eta\to\infty}\frac{\log \vec{u}_*(\eta)}{\eta}=\lambda \quad  \Longrightarrow \quad \vec{u}_*(\eta) \sim e^{\lambda\eta} \, ,
\end{equation}
with $\lambda$ the real part of one of the eigenvalues of $A^\infty_{\ell}$. Now one can easily check that for $\ell > 0$, the constant matrix $A^\infty_{\ell}$ has two positive and two negative eigenvalues:
\begin{equation}
\lambda^1_{\pm}= \pm \sqrt{\frac{v_1}{v_2}\,\ell(\ell+1)} \;,\qquad \lambda^2_{\pm}=\frac12\Bigl(1 \pm\sqrt{1+4\,\frac{v_1}{v_2}\,\ell(\ell+1)}\Bigr) \, ,
\end{equation} 
 and corresponding eigenvectors:
 \begin{equation}
 \vec{u}{\,}^1_\pm = \begin{pmatrix}
 \lambda^1_\pm \\ 0\\0\\ 1
 \end{pmatrix} \;,\qquad  \vec{u}{\,}^2_\pm = \begin{pmatrix}
 0 \\ \lambda^2_\pm \\ 1\\0 
 \end{pmatrix}\ .
 \end{equation}
The general asymptotic solution is then the linear combination:
\begin{equation}
\vec{u}_*(\eta) = c_1\, e^{\lambda^1_+\eta}\, \vec{u}{\,}^1_++ c_2\,e^{\lambda^1_-\eta}\, \vec{u}{\,}^1_-  + c_3\, e^{\lambda^2_+\eta}\, \vec{u}{\,}^2_+ + c_4\,  e^{\lambda^2_-\eta}\, \vec{u}{\,}^2_- \, ,
\end{equation}
where the $c_i$ are constants. From the above, one can read the asymptotic behavior of every field component. For instance we have:
\begin{equation}
\sigma(\eta) \sim c_1 \lambda^1_+\, e^{\lambda^1_+\eta} + \mathcal{O}(e^{\lambda^1_-\eta}) \, .
\end{equation}
Hence the radial mode $\sigma(\eta)$ does not have the required asymptotic behavior \eqref{eq:norm-bc-vec-bos}. Identical results can be obtained for the radial functions $\tilde{W}_{\mu}(\eta)$, so again we find no non-trivial solution to \eqref{eq:diff-eq-sys} satisfying our normalizable boundary conditions.

Note that these considerations are in principle valid for all values of $n$ and $\ell$, so we needed not split the discussion above. However, in the special cases we first discussed one can see in a more obvious way how the boundary and smoothness conditions kill all possible solutions, giving further credibility to our general result.

\subsection{Solving the cokernel equations}
\label{app:cokernel-modes}

To analyze the cokernel equations \eqref{eq:cokerneq-sigma} and \eqref{eq:cokerneq-W}, we make use of the mode expansion along scalar spherical harmonics discussed in Section \ref{sec:mod-exp-bc} for the ghost fields $c$ and $\hat{b}$, as well as for the twisted fermions $\lambda^{ij}$. According to \eqref{eq:scal-mod-exp}, we simply have to track the charge under SU(2)$_\mathrm{R}$ which shifts the $n$ eigenvalue of $\widehat{Q}^2$ by $-1$ for $\lambda^{11}$ and by $+1$ for $\lambda^{22}$. So we use the mode decomposition
\begin{align}
\label{eq:ferm_decomp}
c = &c(\eta)\,e^{\mathrm{i}n\tau}\,Y_\ell{}^m \, , \quad \hat{b} = \hat{b}(\eta)\,e^{\mathrm{i}n\tau}\,Y_\ell{}^m \, , \quad \lambda^{12} = \lambda^{12}(\eta)\,e^{\mathrm{i}n\tau}\,Y_\ell{}^m \, , \nonumber \\
&\lambda^{11} = \lambda^{11}(\eta)\,e^{\mathrm{i}(n-1)\tau}\,Y_\ell{}^m \, , \quad \lambda^{22} = \lambda^{22}(\eta)\,e^{\mathrm{i}(n+1)\tau}\,Y_\ell{}^m \, .
\end{align}
With this decomposition, \eqref{eq:cokerneq-sigma} gives
\begin{align}
\label{eq:cokerneq-sigma-mod}
0 =&\; \Bigl[\lambda(\eta)' + \frac{2}{\cosh\eta\sinh\eta}\,\lambda(\eta) - \sqrt{\frac{v_1}{v_2}}\,\cot\theta\,\mathcal{I}(\eta) + \frac{1}{\sqrt{v_1}\sinh\eta}\,\Delta^{(n,\ell)}c(\eta)\Bigr]\,e^{\mathrm{i}n\tau}\,Y_\ell{}^m \nonumber \\
& - \sqrt{\frac{v_1}{v_2}}\,\Bigl[\mathcal{I}(\eta)\,\partial_\theta Y_\ell{}^m + \mathcal{R}(\eta)\,\frac{1}{\sin\theta}\,\partial_\varphi Y_\ell{}^m\Bigr]\,e^{\mathrm{i}n\tau} 
\end{align}
where we introduced the auxiliary radial functions
\begin{equation}
\label{eq:mode-natural}
\lambda(\eta) := 2\mathrm{i}\,\lambda^{12}(\eta) \, , \quad \mathcal{R}(\eta) := \lambda^{11}(\eta) + \lambda^{22}(\eta) \, , \quad \mathcal{I}(\eta) := \mathrm{i}\,\lambda^{11}(\eta) - \mathrm{i}\,\lambda^{22}(\eta) \, .
\end{equation}
Notice that these functions are also the natural variables appearing in the index theorem computation at the level of the symbol of the operator $D_{10}^\mathrm{vec}$, see \eqref{eq:symb-natural}. To analyze \eqref{eq:cokerneq-W}, we use the explicit expressions for the bilinears $K^{\mu\nu}_{ij}$ and examine each $\mu$ component separately. The $\mu = \tau$ component gives
\begin{align}
\label{eq:cokerneq-W-tau}
0 =&\; \Bigl[\lambda(\eta)' - 2\tanh\eta\,\lambda(\eta) + \frac{\mathrm{i}\,n}{4\sinh\eta}\,\hat{b}(\eta) - \sqrt{\frac{v_1}{v_2}}\,\cot\theta\,\mathcal{I}(\eta) - \frac{\sinh\eta}{\sqrt{v_1}}\,\Delta^{(n,\ell)}c(\eta) \Bigr]\,e^{\mathrm{i}n\tau}\,Y_\ell{}^m \nonumber \\
& - \sqrt{\frac{v_1}{v_2}}\,\Bigl[\mathcal{I}(\eta)\,\partial_\theta Y_\ell{}^m + \mathcal{R}(\eta)\,\frac{1}{\sin\theta}\,\partial_\varphi Y_\ell{}^m\Bigr]\,e^{\mathrm{i}n\tau} \, .
\end{align}
The $\mu = \eta$ component gives
\begin{align}
0 =&\; \Bigl[\hat{b}(\eta)' - \frac{4\mathrm{i}\,n}{\sinh\eta}\,\lambda(\eta) - 4\cosh\eta\,\sqrt{\frac{v_1}{v_2}}\,\cot\theta\,\mathcal{R}(\eta)\Bigr]\,e^{\mathrm{i}n\tau}\,Y_\ell{}^m \nonumber \\
& - 4\cosh\eta\,\sqrt{\frac{v_1}{v_2}}\,\Bigl[\mathcal{R}(\eta)\,\partial_\theta Y_\ell{}^m - \mathcal{I}(\eta)\,\frac{1}{\sin\theta}\,\partial_\varphi Y_\ell{}^m\Bigr]\,e^{\mathrm{i}n\tau} \, .
\end{align}
The $\mu = \theta$ component gives
\begin{align}
0 =&\; \Bigl[\mathcal{R}(\eta)' + \frac{1}{\sinh\eta\cosh\eta}\,\mathcal{R}(\eta) + \frac{\mathrm{i}\,n}{\cosh\eta\sinh\eta}\,\mathcal{I}(\eta)\Bigr]\,e^{\mathrm{i}n\tau}\,Y_\ell{}^m \nonumber \\
& + \sqrt{\frac{v_1}{v_2}}\,\Bigl[\frac{1}{4\cosh\eta}\,\hat{b}(\eta)\,\partial_\theta Y_\ell{}^m + \lambda(\eta)\,\frac{1}{\sin\theta}\,\partial_\varphi Y_\ell{}^m \Bigr]\,e^{\mathrm{i}n\tau} \, .
\end{align}
Lastly the $\mu = \varphi$ component gives
\begin{align}
0 =&\; \Bigl[\mathcal{I}(\eta)' + \frac{1}{\sinh\eta\cosh\eta}\,\mathcal{I}(\eta) -\frac{\mathrm{i}\,n}{\cosh\eta\sinh\eta}\,\mathcal{R}(\eta)\Bigr]\,e^{\mathrm{i}n\tau}\,Y_\ell{}^m \nonumber \\
& + \sqrt{\frac{v_1}{v_2}}\,\Bigl[\lambda(\eta)\,\partial_\theta Y_\ell{}^m - \frac{1}{4\cosh\eta}\,\hat{b}(\eta)\,\frac{1}{\sin\theta}\,\partial_\varphi Y_\ell{}^m \Bigr]\,e^{\mathrm{i}n\tau} \, .
\end{align}
We now recall that, when $m \neq \ell$, the spherical harmonics and their derivatives provide an orthonormal basis for expanding fields on $S^2$. Therefore in this case, the above equations immediately imply
\begin{equation}
\lambda(\eta) = \hat{b}(\eta) = \mathcal{R}(\eta) = \mathcal{I}(\eta) = 0 \, .
\end{equation}
Then \eqref{eq:cokerneq-sigma-mod} reduces to
\begin{equation}
\Delta^{(n,\ell\neq0)}c(\eta) = 0 \, .
\end{equation}
As we saw in our analysis of the kernel equations in Appendix \ref{app:kernel-modes}, our choice of boundary conditions then force $c(\eta)$ to vanish. Hence, when $m \neq \ell$, we find that the only solution in the cokernel of $D_{10}^\mathrm{vec}$ is the trivial solution. With a little extra care one can also see easily that even in the case $m = \ell \neq 0$ we reach the same conclusion, based on the explicit form of the equations that can be used against each other to eliminate the variables one by one.

It remains to discuss the case $\ell = m = 0$. In this case, the spherical harmonics are constant and their derivatives vanish. After using \eqref{eq:cokerneq-sigma-mod} to eliminate $\Delta^{(n,0)}c(\eta)$ from \eqref{eq:cokerneq-W-tau}, we are left with a system of coupled first-order ODEs on the radial modes:
\begin{equation}
\label{eq:co-diff-eq-sys}
\vec{v}(\eta)' = B_n(\eta) \cdot \vec{v}(\eta) \, , 
\end{equation}
with
\begin{equation}
\vec{v}(\eta) = \Bigl(\lambda(\eta) \; , \; \hat{b}(\eta) \; , \; \mathcal{R}(\eta) \; , \; \mathcal{I}(\eta)\Bigr)^\mathrm{T} \, , 
\end{equation}
and
\begin{equation}
B_n(\eta) = \begin{pmatrix} 0 & -\frac{\mathrm{i}n}{4\sinh\eta\cosh^2\eta} & 0 & \sqrt{\frac{v_1}{v_2}}\cot\theta \\ \frac{4\mathrm{i}\,n}{\sinh\eta} & 0 & 4\,\sqrt{\frac{v_1}{v_2}}\cosh\eta\,\cot\theta & 0 \\ 0 & 0 & -\frac{1}{\sinh\eta\cosh\eta} & -\frac{\mathrm{i}n}{\sinh\eta\cosh\eta} \\ 0 & 0 & \frac{\mathrm{i}n}{\sinh\eta\cosh\eta} & -\frac{1}{\sinh\eta\cosh\eta} \end{pmatrix} \, . \vspace{2mm}
\end{equation}
In this form, we see that for any $n \in \mathbb{Z}$ the modes $\mathcal{R}(\eta)$ and $\mathcal{I}(\eta)$ decouple. The solutions to the decoupled system are given by
\begin{align}
\mathcal{R}(\eta) =&\; \frac{1}{2}\,\Bigl[(f_1 + \mathrm{i}\,\textnormal{sign}(n)\,f_2)\tanh^{|n|-1}\eta + (f_1 - \mathrm{i}\,\textnormal{sign}(n)\,f_2)\tanh^{-|n|-1}\eta\,\Bigr] \, , \\
\mathcal{I}(\eta) =&\; \frac{\mathrm{i}}{2}\,\Bigl[(f_1 + \mathrm{i}\,\textnormal{sign}(n)\,f_2)\tanh^{|n|-1}\eta - (f_1 - \mathrm{i}\,\textnormal{sign}(n)\,f_2)\tanh^{-|n|-1}\eta\,\Bigr]\,\textnormal{sign}(n) \nonumber \, ,
\end{align}
where $f_1$ and $f_2$ are the integration constants. To ensure smoothness of $\lambda^{11}(\eta)$ and $\lambda^{22}(\eta)$ at the origin according to \eqref{eq:smooth-vec-ferm}, we must set the constants $f_1 = f_2 = 0$. As a consequence, the radial modes $\mathcal{R}(\eta)$ and $\mathcal{I}(\eta)$ drop out of the system \eqref{eq:co-diff-eq-sys} and what remains is a system of two coupled first-order ODEs on $\lambda(\eta)$ and $\hat{b}(\eta)$. We know discuss its potential solutions depending on whether the quantum number $n$ vanishes or not.\\

\underline{\textbf{The case} $\mathbf{n\neq 0}$}:\\

When $n\neq0$, the system of ODEs on $\lambda(\eta)$ and $\hat{b}(\eta)$ can be written as
\begin{equation}
\lambda(\eta) = \frac{\sinh\eta}{4\mathrm{i}n}\,\hat{b}(\eta)'  \, , \quad \hat{b}(\eta)'' + \coth\eta\;\hat{b}(\eta)' - \frac{n^2}{\sinh^2\eta\cosh^2\eta}\,\hat{b}(\eta) = 0 \, .
\end{equation}
Note that the equation for $\hat{b}(\eta)$ is the same as the one for $\sigma(\eta)$ in \eqref{eq:sys-n0}. The solution for $\hat{b}(\eta)$ takes the same form,
\begin{equation}
\label{eq:bsol-nneq0}
\hat{b}(\eta) = f_3\tanh^{|n|}\eta\;{}_2F_1\Bigl(\frac12|n| + \frac14\bigl(1 - \sqrt{1 + 4n^2}\bigr),\frac12|n| + \frac14\bigl(1 + \sqrt{1 + 4n^2}\bigr);1+|n|;\tanh^2\eta\Bigr) \, , 
\end{equation}
but now our boundary conditions \eqref{eq:susy-bc-ghost} do not require $f_3$ to vanish since $\hat{b}(\eta)$ is allowed to go to a constant when $\eta \rightarrow \infty$. Then, coming back to \eqref{eq:cokerneq-sigma-mod}, we have an inhomogeneous second-order differential equation on the ghost radial mode $c(\eta)$,
\begin{equation}
\Delta^{(n,0)}c(\eta) = \mathrm{i}\,\sqrt{v_1}\,\Bigl(\frac{1}{2n}\tanh\eta\;\hat{b}(\eta)' + \frac{n}{4\cosh^2\eta}\,\hat{b}(\eta)\Bigr) \, ,
\end{equation}
with $\hat{b}(\eta)$ given by \eqref{eq:bsol-nneq0}. It is straightforward to check that a particular solution to this equation is given by
\begin{equation}
c(\eta) = -\mathrm{i}\,\frac{\sqrt{v_1}}{4n}\,\hat{b}(\eta) \, .
\end{equation}
Since the homogeneous equation $\Delta^{(n,0)}c(\eta) = 0$ has no non-trivial smooth and asymptotically well-behaved solution for $n \neq 0$, this exhausts the possible set of solutions.

In conclusion, we have shown that the cokernel of $D_{10}^\mathrm{vec}$ is non-empty for $n \neq 0$ and $\ell = 0$. It has dimension one since we found a one-parameter family of smooth and asymptotically well-behaved solutions:\footnote{Here we use \eqref{eq:hatb} to express our results on the anti-ghost field $b(\eta)$ directly.}
\begin{align}
&\lambda^{11}(\eta) = \lambda^{22}(\eta) = 0 \, , \;\; \lambda^{12}(\eta) = -\frac{\sinh\eta}{4n}\,b(\eta)' \, , \;\; c(\eta) = -\mathrm{i}\,\frac{\sqrt{v_1}}{2n}\,b(\eta) \, , \\
&b(\eta) = \frac12\,f_3\tanh^{|n|}\eta\;{}_2F_1\Bigl(\frac12|n| + \frac14\bigl(1 - \sqrt{1 + 4n^2}\bigr),\frac12|n| + \frac14\bigl(1 + \sqrt{1 + 4n^2}\bigr);1+|n|;\tanh^2\eta\Bigr) \, , \nonumber \vspace{2mm}
\end{align}
where $f_3$ is an arbitrary constant.\\

\underline{\textbf{The case} $\mathbf{n = 0}$}:\\

When $n = 0$, the system \eqref{eq:co-diff-eq-sys} with $\mathcal{R}(\eta) = \mathcal{I}(\eta) = 0$ collapses, and the solution is
\begin{equation}
\hat{b}(\eta) = f_4 \, , \qquad \lambda(\eta) = f_5 \, ,
\end{equation}
with $f_4$ and $f_5$ are the integration constants. For $\lambda(\eta)$ to be smooth, we must set $f_5 = 0$ according to \eqref{eq:smooth-vec-ferm}. Then coming back to \eqref{eq:cokerneq-sigma-mod}, we have the last equation for the radial mode of the ghost field,
\begin{equation}
\Delta^{(0,0)}c(\eta) = 0 \, .
\end{equation}
We already saw in the kernel analysis of Appendix \ref{app:kernel-modes} that the solution compatible with our boundary conditions is
\begin{equation}
c(\eta) = f_6 \, ,
\end{equation}
with $f_6$ a constant. Therefore, when $(n,\ell) = (0,0)$, we find that the only solutions are two constant modes for the ghost fields,
\begin{equation}
\label{eq:cokernel-zm}
\lambda^{12}(\eta) = \lambda^{11}(\eta) = \lambda^{22}(\eta) = 0 \, , \quad b(\eta) = f_4 \, , \;\; c(\eta) = f_6 \, . 
\end{equation}
This concludes the analysis of the kernel and cokernel differential equations.

\providecommand{\href}[2]{#2}

\end{document}